\documentclass{appolb}
\usepackage{graphicx}
\usepackage{braket}
\usepackage{bbold}
\usepackage{amsmath,amssymb,amsthm,mathtools,bm}
\usepackage{physics}
\usepackage{bbm}
\usepackage{hyperref}
\usepackage[numbers, sort&compress]{natbib}
\newcommand{\be}{\begin{equation}}
\newcommand{\ee}{\end{equation}}

\begin{document}
\title{The Maximal Entanglement Limit\\ \\
in Statistical and High Energy Physics
\thanks{Lectures at the ${65^{\rm th}}$ Jubilee Cracow School of Theoretical Physics, Zakopane, Tatra mountains, Poland, June 14-21, 2025}\\%
}
\author{Dmitri E. Kharzeev
\address{Center for Nuclear Theory, Department of Physics and Astronomy,\\
Stony Brook University, Stony Brook, New York 11794-3800, USA}
\\[3mm]
\address{Energy and Photon Sciences Directorate,\\ Condensed Matter and Materials Sciences Division,\\
Brookhaven National Laboratory, Upton, New York 11973-5000, USA}
}
\maketitle
\begin{abstract}
These lectures advocate the idea that quantum entanglement provides a unifying foundation for both statistical physics and high-energy interactions. I argue that, at sufficiently long times or high energies, most quantum systems approach a Maximal Entanglement Limit (MEL) in which phases of quantum states become unobservable, reduced density matrices acquire a thermal form, and probabilistic descriptions emerge without invoking ergodicity or classical randomness. Within this framework, the emergence of probabilistic parton model, thermalization in the break-up of confining strings and in high-energy collisions, and the universal small $x$ behavior of structure functions arise as direct consequences of entanglement and geometry of high-dimensional Hilbert space. 
\end{abstract}

\newpage  
\section{Introduction}
Why do we age? Why does time appear to have a preferred direction, even though the fundamental laws governing atoms, nuclei, and elementary particles are reversible? How does irreversible macroscopic behavior, and probabilistic statistical physics and the parton model, emerge from the unitary dynamics of quantum mechanics? 

The ubiquity of probabilistic descriptions is indeed one of the most striking features shared by statistical physics and high--energy scattering. Thermodynamics is formulated in terms of ensembles, temperatures, and entropy, while the parton model describes hadrons as collections of quasi--free constituents with probability distributions (structure functions). Yet, at a fundamental level, both domains are governed by quantum mechanics, where the dynamics is unitary and coherent. This raises a fundamental question: how do probabilistic laws and thermal behavior emerge from the quantum dynamics of pure states?

In statistical physics, this question is traditionally answered by invoking ergodicity and coarse graining in phase space. In high--energy physics, the parton model is  justified by appealing to time dilation on the light cone. As we will argue, these explanations obscure a common underlying mechanism: the central role of quantum entanglement. Over the past decades, it has become increasingly clear that entanglement, rather than classical randomness, provides the natural language for understanding thermalization and emergence of probabilistic descriptions in isolated quantum systems.

The central theme of these lectures is the notion of a \emph{Maximal Entanglement Limit} (MEL). We argue that at sufficiently high energies, or after sufficiently long unitary quantum evolution, interactions drive systems toward states in which all information except a small set of conserved quantities is stored nonlocally in entanglement. In this limit, phases of quantum states become unobservable, reduced density matrices acquire thermal (maximum--entropy) forms, and probabilistic descriptions emerge without invoking ergodicity or stochastic assumptions. Entropy growth, thermal features of hadron production in the break-up of confining strings, fast thermalization in high-energy collisions, and the universality of structure functions at small Bjorken $x$ are then understood as direct consequences of entanglement generation and the tracing over unobserved degrees of freedom.

These lectures develop the MEL perspective across a range of different topics. We begin with general concepts of typicality in high-dimensional Hilbert spaces, reduced density matrices, and the number--phase uncertainty relation, and show how they lead to probabilistic descriptions. We then turn to high--energy scattering, where Lorentz time dilation and scale--invariant branching dynamics provide natural mechanisms for the loss of phase information. The emergence of geometric and exponential multiplicity distributions, linear growth of entanglement entropy with rapidity, and the success -- as well as the limitations -- of the parton model are reinterpreted through the MEL. Finally, we present first--principles quantum simulations of massive Schwinger model that explicitly demonstrate how thermal reduced density matrices and maximal entanglement arise from unitary evolution.
\smallskip

By viewing statistical physics and high--energy QCD through the prism of quantum entanglement, the Maximal Entanglement Limit offers a common foundation for phenomena that have long been treated separately. We hope that this perspective will not only clarify some of the old puzzles, but also suggest new ways to think about entropy, information, and universality in relativistic quantum systems.
\bigskip

The subject of quantum entanglement across statistical physics and complex quantum dynamics is vast and rapidly evolving. Lack of space -- and, more importantly, lack of expertise on the part of the author -- does not allow a comprehensive discussion of all its facets. In particular, many powerful and complementary approaches to quantum complexity, chaos, and entanglement lie beyond the scope of these lectures.
\vskip0.2cm

To partially compensate for this limitation, and to guide the interested reader, we list below a small selection of representative reviews and pedagogical articles covering closely related topics. 
These references place our discussion of connections between high--energy QCD and statistical physics in a broader landscape.
\begin{itemize}
\item {\it{Quantum chaos and thermalization:}}
  Reviews of eigenstate thermalization, quantum chaos, and entanglement growth in many--body systems can be found in
  \cite{DAlessio2016,Borgonovi:2016nrx,Deutsch2018}.

\item {\it{Operator growth and OTOCs:}} 
  For introductions to out--of--time--ordered correlators and quantum information scrambling, see
  \cite{Maldacena2016,Roberts2016}.

\item {\it Krylov complexity and computational measures:}  
  A modern perspective on Krylov computational complexity and its role in quantum dynamics is reviewed in
  \cite{Parker2019,Brandao:2019sgy}.

\item {\it Holography, black holes, and entanglement:}  
  The relation between entanglement, spacetime geometry, and black hole physics is reviewed in
  \cite{Nishioka:2009un,Swingle:2009bg,VanRaamsdonk2010,Harlow2016}.
  
  \item {\it Quantum simulations in high-energy and nuclear physics:}  
  We will discuss the quantum simulations of jet fragmentation and string break-up below, but the field is much broader, as reviewed in \cite{Bauer:2022hpo,Bauer:2023qgm}.
  
\end{itemize}

The different approaches reviewed in these papers touch upon a common theme that will be the main subject of these lectures: entanglement and the geometry of Hilbert space play a central role in organizing quantum dynamics, from many-body systems to gravity and high-energy and nuclear physics.

\section{Classical ensembles and quantum mechanics}\label{qm}

In his groundbreaking work, Ludwig Boltzmann laid the conceptual basis of statistical physics by formulating an assumption  that in thermal equilibrium all microscopic states compatible with macroscopic constraints occur with equal probability. As Boltzmann famously wrote, ``Es ist klar, da\ss\ im W\"armegleichgewicht alle m\"oglichen Zust\"ande des Systems \dots\ gleich wahrscheinlich sind''{\footnote{It is clear that, in thermal equilibrium, all possible states of the system $...$ are equally likely.}} \cite{Boltzmann1877}. This statement was not derived from microscopic dynamics; it was instead formulated as the foundational postulate. Boltzmann then devoted much of his scientific life to the ambitious program of justifying this assumption. His efforts led to seminal concepts such as ergodicity and the celebrated ${\rm H}$-theorem, intended to explain the irreversible growth of entropy from reversible dynamics \cite{Boltzmann-translate}. Yet, as is now well appreciated, these arguments ultimately rely on assumptions about typicality in phase space that cannot be rigorously established within classical mechanics alone.

 As I will argue in these lectures, modern developments in quantum information theory allow us to revisit Boltzmann's dilemma from a radically different perspective. Rather than invoking ergodic motion in an abstract phase space, one appeals to the geometry of high-dimensional Hilbert space itself. {\it Typical} pure quantum states, drawn from the Haar measure, are overwhelmingly entangled, and their small subsystems are described by reduced density matrices that are indistinguishable from thermal ensembles. In this view, the principle of equal {\it a priori} probabilities is no longer an assumption, but an emergent consequence of quantum typicality and entanglement -- providing a resolution of the problem that Boltzmann first posed more than a century ago.
\smallskip

The first step in this direction was made by John von Neumann in 1929.  His seminal paper ``Beweis des Ergodensatzes und des H-Theorems in der neuen Mechanik"\footnote{``Proof of the ergodic theorem and the H-theorem in new (quantum) mechanics."}\cite{vonNeumann1929} marked a decisive turning point in the foundations of statistical physics. In this work, von Neumann demonstrated that quantum mechanics provides a fundamentally new resolution of the long-standing puzzles associated with the emergence of thermodynamics from reversible microscopic laws. 
\smallskip

Unlike classical statistical mechanics, where ergodicity must be postulated and remains notoriously difficult to justify, von Neumann argued that typical pure quantum states of large isolated systems already exhibit thermal behavior when restricted to macroscopic observables. His analysis introduced the notion of typicality: for almost all states within a narrow energy shell, expectation values of coarse-grained observables coincide with those computed in the microcanonical ensemble. In this sense, thermodynamic behavior emerges not from time averaging over trajectories, but from the structure of the quantum Hilbert space itself. 
Moreover, von Neumann's quantum version of the H-theorem clarified how irreversible entropy increase arises from unitary dynamics once one accounts for the inevitable coarse-graining associated with macroscopic measurements. 
\smallskip

Following von Neumann's pioneering work, a long sequence of developments clarified and sharpened his original insights, progressively revealing entanglement as the central mechanism behind quantum thermalization. Early advances formalized the idea of typicality in large Hilbert spaces, showing that for generic pure states within a narrow energy window, small subsystems are overwhelmingly likely to be described by thermal density matrices \cite{Goldstein:2005aib,Popescu:2006rhr}. These results, often referred to as canonical or microcanonical typicality, made explicit that thermal behavior emerges from entanglement between a subsystem and its complement, rather than from dynamical chaos or ensemble averaging. 
\smallskip

In parallel, the study of quantum chaos and random matrix theory provided a dynamical underpinning for typicality, culminating in the Eigenstate Thermalization Hypothesis (ETH), which posits that individual energy eigenstates of nonintegrable systems already encode thermal expectation values for local observables \cite{Deutsch:1991msp,Srednicki:1994mfb,Rigol:2007juv}. From this perspective, thermalization reflects the highly entangled structure of many-body eigenstates themselves. 
\vskip0.2cm

Quantum information theory subsequently supplied an appropriate language for these ideas, emphasizing reduced density matrices, entanglement entropy, and entanglement spectrum as the natural descendants of von Neumann's original arguments. At the same time, integrable systems demonstrated sharp and physically illuminating exceptions, where integrability prevents full thermalization \cite{Nandkishore:2014kca}. Together, these developments firmly established entanglement -- not ergodicity or subjective probability -- as the fundamental link between unitary quantum dynamics and thermodynamic behavior.
\smallskip

When a system \( S \) interacts with a larger environment \( E \), the joint state \( \rho_{SE} \) evolves unitarily and remains pure, yet the reduced state of the subsystem,
\begin{equation}
\rho_S = \Tr_E \rho_{SE},
\end{equation}
is generically mixed. This loss of purity is not a consequence of stochasticity, but rather a direct manifestation of entanglement between \( S \) and \( E \). The associated entanglement entropy,
\begin{equation}
S(\rho_S) = -\Tr\!\left(\rho_S \ln \rho_S\right),
\end{equation}
provides a quantitative measure of the information delocalized into environmental degrees of freedom and thus supplies a precise link between microscopic unitary dynamics and macroscopic thermodynamic behavior. 
\smallskip

In this framework, the entropy that appears in equilibrium and nonequilibrium statistical physics is identified with the entanglement structure of typical quantum states. Ensemble descriptions and thermal expectation values then emerge as generic properties of highly entangled pure states, rather than as fundamental probabilistic postulates. 
By grounding statistical mechanics in quantum entanglement, we arrive at a fully quantum-mechanical foundation for irreversibility and thermalization. To see this explicitly, let us first briefly review the postulates of classical statistical physics.

\section{Classical Statistical Physics and Ergodicity}

Classical statistical mechanics is built upon a set of foundational assumptions that allow one to describe the macroscopic properties of many-body systems in terms of probabilistic distributions over phase space. Despite the fact that classical mechanics is governed by deterministic Hamiltonian evolution, these postulates introduce ensemble descriptions that effectively encode our lack of knowledge of microscopic details and enable the emergence of thermodynamic behavior. In this section we briefly review the basic constructs on which the theory is built.

\subsection{Microcanonical ensemble}\label{microcan}

The most fundamental statistical description of an isolated classical system with fixed total energy $E$ is the \emph{microcanonical ensemble}. In this ensemble, all phase-space points $z$ satisfying the energy constraint $H(z)=E$ are assumed to be equally likely. The corresponding probability density on phase space is
\begin{equation}\label{en-shell}
  \rho_{\mathrm{mc}}(z) = \frac{\delta(H(z)-E)}{\Omega(E)} ,
\end{equation}
where $\Omega(E)$ denotes the density of states at energy $E$, ensuring proper normalization. This postulate embodies the idea that, in equilibrium, the system explores all energetically allowed microstates with equal weight, reflecting maximal ignorance about microstructure consistent with the fixed energy.

\subsection{Canonical ensemble from entropy maximization}

When a system is in contact with a heat bath at fixed temperature $T \equiv 1/\beta$, the energy of the subsystem is no longer fixed but fluctuates due to exchanges with the environment. The appropriate equilibrium distribution is then the \emph{canonical ensemble}. 
\smallskip

Maximizing Gibbs entropy $S[\rho]= -\!\int \rho \ln \rho\,\dd\Gamma$ at fixed $\langle H\rangle=E$ yields
\begin{equation}
\rho_{\mathrm{can}}(z) = \frac{e^{-\beta H(z)}}{Z(\beta)},\qquad
Z(\beta) \equiv \int_\Gamma e^{-\beta H(z)}\,\dd\Gamma,
\end{equation}
with inverse temperature $\beta$ set by $E=-\partial_{\beta}\ln Z$.
This distribution captures the statistical weight of microstates at a given temperature and underlies the thermodynamic relations connecting energy, entropy, and free energy.

\smallskip

For $H=H_A+H_B$ with negligible interaction, the partition function factorizes:
\begin{equation}
Z_{A\cup B}(\beta) = Z_A(\beta)\,Z_B(\beta),\qquad
\rho_{\mathrm{can}}^{A\cup B}(z_A,z_B) = \rho_{\mathrm{can}}^{A}(z_A)\,\rho_{\mathrm{can}}^{B}(z_B) .
\end{equation}
The entropy of the entire system is then given by the sum of entropies of subsystems: $S=S_A + S_B$\footnote{Note that, in contrast to classical systems, for a pure quantum state $S=0$ whereas the entanglement entropies of the subsystems in general do not vanish, $S_A, S_B \geq 0$.}.

\subsection{Ergodic hypothesis}

A crucial conceptual ingredient often invoked to justify the connection between microscopic dynamics and ensemble averages is the \emph{ergodic hypothesis}, originally formulated by Ludwig Boltzmann. It asserts that, for sufficiently long times, the time average of any observable $A(z)$ along a single dynamical trajectory equals its ensemble average in the microcanonical distribution:
\begin{equation}
  \overline{A} = \langle A \rangle_{\mathrm{mc}} .
\end{equation}
Although rigorous proofs exist only for special systems, and many physical systems are known not to be strictly ergodic, this hypothesis encapsulates the idea that dynamical motion allows the system to explore its entire phase space, 
and the time spent by the system in some region of phase space is proportional to the volume of this region. Therefore, all accessible microstates are assumed equiprobable over a sufficently long period of time. 
The ergodic hypothesis thus provides a conceptual bridge between deterministic dynamics and the probabilistic assumptions of equilibrium statistical mechanics.

\section{Maximally Entangled States from Geometry of Hilbert Space}\label{geomh}

In quantum mechanics, entanglement plays a central role in understanding how thermodynamic behavior can arise from purely unitary and information-preserving dynamics. When a large quantum system is partitioned into a subsystem $S$ and an environment $E$, the correlations between them give rise to effective mixedness and entropy at the level of $S$, even when the global state is strictly pure. This section reviews key information-theoretic concepts that connect entanglement to statistical behavior.

\subsection{Reduced density matrices and entanglement entropy}

Consider a global pure state $\ket{\Psi}$ defined on a bipartite Hilbert space $\mathcal{H} = \mathcal{H}_S \otimes \mathcal{H}_E$. The state of the subsystem $S$ is described by the reduced density matrix\footnote{The formalism of density matrices was introduced independently by L.D. Landau \cite{Landau:1927gxg} and J. von Neumann \cite{vonNeumann1929den}.} $\rho_S = \Tr_E \ket{\Psi}\bra{\Psi},$
obtained by tracing out the degrees of freedom of the environment. Even though $\ket{\Psi}$ is a pure state, $\rho_S$ is in general mixed whenever $S$ and $E$ are entangled. This mixedness encapsulates the information about $S$ that is inaccessible without access to $E$.


The von Neumann entropy $S(\rho_S) = - \Tr(\rho_S \ln \rho_S)$
quantifies the degree of entanglement between $S$ and $E$. For any global pure state $\ket{\Psi}$, the subsystem entropies satisfy
\begin{equation}
  S(\rho_S) = S(\rho_E),
\end{equation}
reflecting the fact that entanglement entropy measures shared quantum correlations rather than local disorder. Entanglement entropy thus provides a natural notion of thermodynamic entropy for subsystems of large quantum systems evolving unitarily.

\subsection{Typical states and maximal entanglement} \label{typ}

A remarkable result in quantum information theory is that for a {\it typical} pure state of a large composite system, the reduced state of a small subsystem is extremely close to the maximally mixed state. More precisely, Page's theorem \cite{Page:1993df} (see also \cite{Lubkin:1978nch,Lloyd:1988cn}) states that for a Haar-random\footnote{Haar measure was introduced by Alfr\'ed Haar in 1933 as an invariant volume assigned to locally compact topological groups. As a simple example, Haar measure of the $U(1)$ group is $d \mu = d\theta/2\pi$.} pure state $\ket{\Psi}$ on $\mathcal{H}_S \otimes \mathcal{H}_E$ with $d_E \gg d_S$, one finds
\begin{equation}\label{eqprob}
  \rho_S \approx \frac{\mathbb{1}}{d_S}.
\end{equation}
This demonstrates that maximal entanglement is overwhelmingly likely in high-dimensional Hilbert spaces and provides a mechanism by which thermal-like states can emerge from pure quantum states without invoking ensembles or stochastic assumptions. In the following Sections \ref{rand} and \ref{page} we will sketch the proof of this theorem; however let us first discuss its geometric origin.
\vskip0.2cm

The idea that thermodynamic behavior should be understood as a \emph{typical} property of quantum states can be traced back directly to von Neumann's 1929 analysis \cite{vonNeumann1929} discussed above in Section \ref{qm}. Von Neumann already emphasized that for macroscopic systems, almost all pure states within a narrow energy window yield the same expectation values for coarse-grained observables, anticipating by several decades the modern notion of typicality. In contemporary language, this statement is formulated using the Haar measure, the unique unitarily invariant probability measure on the unit sphere of a Hilbert space $\mathcal H$, or of a constrained subspace $\mathcal M\subset\mathcal H$ such as a microcanonical energy shell. A property of states $|\psi\rangle\in\mathcal M$ is said to be typical if the fraction of states that violate it vanishes exponentially as $\dim\mathcal M\to\infty$. Typicality is therefore a statement about the geometry of Hilbert space, not a dynamical assumption about the properties of time evolution.
\vskip0.2cm

This perspective is particularly natural in high-energy scattering, where the relevant Hilbert space is the Fock space of multiparticle states. At high energies, Lorentz time dilation render the relative phases between different Fock components effectively unobservable. Tracing over these inaccessible phase degrees of freedom selects a reduced density matrix that samples a typical state within the accessible subspace. As the collision energy increases and the available phase space grows, concentration of measure ensures that the reduced state becomes highly entangled and approaches a universal, nearly thermal form. In this way, the appearance of thermal spectra in high-energy processes can be understood as a direct consequence of typicality and entanglement in large Hilbert spaces, rather than as evidence for conventional equilibration through rescattering. We will discuss this in more detail below.

\subsection{Typical states and random matrices}\label{rand}

In view of importance of the result (\ref{eqprob}), let us sketch its proof. 
Consider a bipartite Hilbert space $\mathcal H=\mathcal H_A\otimes \mathcal H_B$ with
\[
\dim\mathcal H_A=m,\qquad \dim\mathcal H_B=n,\qquad m\le n,
\]
and draw a Haar-random pure state $|\psi\rangle\in\mathcal H$. The reduced density matrix
\[
\rho_A=\Tr_B |\psi\rangle\langle\psi|
\]
is then a random positive matrix of size $m\times m$. This Haar ensemble can be related to a Wishart random matrix ensemble \cite{ZyczkowskiSommers2001,CollinsSniady2006}, named after John Wishart who introduced these matrices in 1928 to generalize the gamma distribution to multiple dimensions.
\smallskip

Wishart matrices arise naturally in the study of reduced density matrices of subsystems of large quantum systems and therefore play a central role in the theory of typical quantum states. A Wishart matrix is a positive-semidefinite matrix of the form
\[
W = X X^\dagger ,
\]
where $X$ is a rectangular matrix whose entries are independent complex Gaussian random variables. Originally introduced in multivariate statistics to describe sample covariance matrices\footnote{The Wishart distribution can be viewed as a multivariate generalization of the $\chi^2$ distribution, representing the distribution of symmetric positive-definite matrices.}, Wishart ensembles enter quantum mechanics whenever a pure state on a bipartite Hilbert space $\mathcal H_A\otimes\mathcal H_B$ is chosen at random with respect to the Haar measure. In this construction, the reduced density matrix $\rho_A=\Tr_B |\psi\rangle\langle\psi|$ can be written as a normalized Wishart matrix,
\be \label{eq:rhoWishart}
\rho_A = \frac{W}{\Tr W}, \qquad W = X X^\dagger ,
\ee
so that the statistical properties of $\rho_A$ are governed by the Wishart unitary ensemble.
\smallskip

The eigenvalues $\{\lambda_i\}$ of $\rho_A$ are distributed according to
\be \label{measure}
P(\{\lambda_i\}) \propto 
\delta\!\left(1-\sum_{i=1}^m \lambda_i\right)
\prod_{i=1}^m \lambda_i^{\,n-m}
\prod_{i<j}(\lambda_i-\lambda_j)^2 ,
\ee
which makes explicit the strong eigenvalue repulsion characteristic of random matrix ensembles. This Vandermonde determinant is mathematically identical to the one first encountered by Wigner in his description of nuclear energy-level statistics \cite{Wigner1951} and later systematized in the Wigner--Dyson ensembles. The key distinction is conceptual: in Wigner's original work, random matrices model complicated nuclear Hamiltonians, whereas in the present context Wishart matrices describe reduced density matrices obtained by tracing out environmental degrees of freedom. Despite this difference, both ensembles share the same underlying unitary invariance, and hence the same universal spectral correlations. In view of the importance of this connection, we discuss it in more detail in Section \ref{Wig}.
\smallskip

From the perspective of typicality, Wishart matrices provide a concrete and calculable realization of concentration of measure. In high dimensions, the eigenvalues of a Wishart matrix cluster near the equiprobable value because concentration of measure and global entropic dominance overwhelm local eigenvalue repulsion, which only enforces small local level spacings without preventing maximal mixing. The induced eigenvalue distribution is thus sharply peaked around $\lambda_i\simeq 1/m$ ($m=N_A$ is the number of eigenvalues), implying that $\rho_A$ is, with overwhelming probability, close to the maximally mixed state. 

\subsection{Maximal entanglement and $2D$ Coulomb gas}

A particularly intuitive way to understand why the eigenvalues of a high--dimensional Wishart matrix cluster near equiprobability is provided by the Coulomb--gas representation of random matrix ensembles \cite{Dean:2006wk}. The interpretation of the Vandermonde factor as Coulomb repulsion follows directly from the logarithmic form of the electrostatic interaction in two dimensions. In random matrix ensembles, the joint probability distribution of eigenvalues contains the factor
\begin{equation}
\prod_{i<j}(\lambda_i-\lambda_j)^2
= \exp\!\left[\,2\sum_{i<j}\ln|\lambda_i-\lambda_j|\,\right].
\end{equation}
When this probability distribution is written in Boltzmann form, $P\propto e^{-\beta H_{\rm eff}}$, the exponent defines an effective Hamiltonian
\begin{equation}
H_{\rm eff} \supset - \sum_{i<j}\ln|\lambda_i-\lambda_j| ,
\end{equation}
up to an overall constant. This is precisely the interaction energy of identical charges confined to a line but interacting via a two--dimensional Coulomb potential, for which the electrostatic potential between charges at positions $\lambda_i$ and $\lambda_j$ is logarithmic,
\begin{equation}
V_{\rm Coulomb}(\lambda_i-\lambda_j)\;\propto\; -\ln|\lambda_i-\lambda_j| .
\end{equation}

The origin of this logarithmic interaction is geometric. Eigenvalues arise from diagonalizing a matrix, and the Jacobian of the transformation from matrix elements to eigenvalues and eigenvectors is given by the Vandermonde determinant. The vanishing of this determinant when $\lambda_i=\lambda_j$ reflects the singularity of the coordinate transformation at degenerate spectra, and in the statistical ensemble this singularity manifests itself as an effective repulsion between eigenvalues. The square of the Vandermonde determinant therefore enforces level repulsion by assigning an infinite energetic cost to coincident eigenvalues.
\smallskip

In the Coulomb--gas analogy, eigenvalues behave as like--charged particles whose mutual repulsion is logarithmic, while external potential terms and normalization constraints provide confinement. Indeed, the additional factor
\begin{equation}
\prod_i \lambda_i^{\,n-m}
\end{equation}
in the joint eigenvalue distribution plays the role of an external confining potential. Writing the probability distribution in Boltzmann form,
\begin{equation}
P(\{\lambda_i\}) \propto e^{-\beta H_{\rm eff}},
\end{equation}
this factor contributes an effective one--body term to the Hamiltonian,
\begin{equation}\label{coul}
H_{\rm eff} \supset - (n-m)\sum_i \ln \lambda_i .
\end{equation}
Thus each eigenvalue experiences a logarithmic potential that diverges as $\lambda_i \to 0$.

In the Coulomb--gas analogy, this term acts as a repulsive potential that repels eigenvalues from the origin -- remember that $m \leq n$. It reflects the fact that the Wishart matrix is positive definite and that configurations with very small eigenvalues occupy a parametrically smaller volume in matrix space. The factor $(n-m)$ is determined by the rectangularity of the underlying random matrix and encodes how strongly the measure suppresses eigenvalues near zero.

Together with the repulsion between eigenvalues and the normalization constraint $\sum_i \lambda_i = 1$, this logarithmic mean-field repulsion stabilizes the Coulomb gas. Eigenvalues are prevented from collapsing onto the center at $\lambda=0$, the normalization constraint does not allow the eigenvalues to separate far from the center, while mutual repulsion prevents degeneracies. The resulting equilibrium configuration is a compact ``droplet", with density distribution fixed by the balance between the repulsive mean-field potential and repulsion between individual eigenvalues. 

In the large--dimension limit $N_B \gg N_A \gg 1$, the typical spacing between eigenvalues scales as
$$
\delta\lambda \sim \frac{1}{N_A\sqrt{N_B}},
$$
so that the spectrum becomes sharply clustered around the equiprobable value $1/N_A$ as the size of the environment increases.
Therefore, the entropic dominance at large dimensions leads to the clustering of eigenvalues near the maximally mixed value $\lambda_i \simeq 1/N_A$ corresponding to the Maximal Entanglement Limit (MEL).
\smallskip

Consequently, the entanglement entropy of a typical pure state is nearly maximal, as quantified by Page's theorem. In this way, Wishart random matrices link Haar-random states, typicality in Hilbert space, and the emergence of thermal behavior in subsystems of isolated quantum systems.

\subsection{Maximal entanglement from typicality}\label{page}

Let us now show that the reduced density matrix $\rho_A$ of the form \eqref{eq:rhoWishart},\eqref{measure} leads to the Boltzmann-like formula for the entanglement entropy $S(\rho_A)$. 

The von Neumann entropy $S(\rho_A)=-\Tr(\rho_A\ln\rho_A)$ can be obtained from Renyi entropies via the replica identity
\begin{equation}
S(\rho_A)= -\left.\frac{\partial}{\partial q}\ln \Tr(\rho_A^q)\right|_{q=1}.
\label{eq:replica}
\end{equation}
Thus we seek $\left\langle \Tr(\rho_A^q)\right\rangle$ over the ensemble \eqref{eq:rhoWishart}--\eqref{measure}. For integer $q\ge 2$, one may evaluate the moment using Wishart averages \cite{CollinsSniady2006,ZyczkowskiSommers2001}. The result can be expressed in closed form in terms of gamma functions; differentiating and analytically continuing to $q\to 1$ yields Page's formula \cite{Page:1993df,FoongKanno1994,Sen1996}:
\begin{equation}
\boxed{
\big\langle S(\rho_A)\big\rangle
=
\sum_{k=n+1}^{mn}\frac{1}{k}
-\frac{m-1}{2n} .
}
\label{eq:PageExact}
\end{equation}
Equivalently, using harmonic numbers $H_\ell=\sum_{k=1}^{\ell}1/k$,
\begin{equation}
\big\langle S(\rho_A)\big\rangle
= H_{mn}-H_n-\frac{m-1}{2n}.
\end{equation}



For $n\gg m$ one expands the harmonic numbers,
\begin{equation}
H_{mn}-H_n=\ln m+\mathcal O\!\left(\frac{1}{n}\right),
\end{equation}
and obtains
\begin{equation}
\boxed{
\big\langle S(\rho_A)\big\rangle
=
\ln m -\frac{m}{2n}+ \mathcal O\!\left(\frac{1}{n^2}\right)
}.
\label{eq:PageAsympt}
\end{equation}
Hence a \emph{typical} Haar-random pure state is so entangled that the reduced state is close to maximally mixed,
$\rho_A\simeq \mathbb I_m/m$, and the subsystem entropy is essentially the maximum $\ln m$. 
\smallskip

Note that this is equivalent to the Boltzmann's formula for the entropy $S = \ln W$ if we identify the number of quantum states $m$ in the Hilbert space of the subsystem $A$ with the number of microstates $W$.
This random-matrix result thus provides a quantitative underpinning for entanglement-driven thermalization:
the appearance of a thermal (mixed) state for $A$ is generic and originates from tracing out $B$ rather than from
any fundamental probabilistic postulate.

\subsection{Maximal entanglement and nuclear level statistics}\label{Wig}

It is worth emphasizing that the induced eigenvalue measure \eqref{measure} has deep historical roots in nuclear physics. Long before its modern applications to quantum information and entanglement, the same mathematical structure first appeared in Eugene Wigner's seminal 1951 work \cite{Wigner1951} on the statistics of nuclear energy levels. In his attempt to describe the complex spectra of heavy nuclei, Wigner proposed modeling the nuclear Hamiltonian as a random matrix drawn from an appropriate symmetry class, leading to the characteristic joint probability distribution for eigenvalues with logarithmic repulsion, $\prod_{i<j}(\lambda_i-\lambda_j)^2$ \cite{Wigner1951,MehtaBook}. This level-repulsion factor, which also appears in \eqref{measure}, reflects the underlying unitary invariance of the ensemble and is a universal signature of quantum complexity. 
\smallskip

In the present context, the same random-matrix statistics governs the spectrum of reduced density matrices obtained by tracing over environmental degrees of freedom. Thus, Page's theorem can be viewed as a direct descendant of Wigner's original insight: randomness at the level of quantum amplitudes leads to universal statistical properties of spectra, whether of nuclear Hamiltonians or of entanglement in multi-dimensional quantum states.

\section{Maximal Entanglement and Thermodynamic Entropy}\label{canons}

We are now ready to discuss how the thermodynamic entropy arises from the maximally entangled state \eqref{eqprob}. For this, we first need to define this state for a ``microcanonical shell" of states with a fixed energy, in accord with the microcanonical approach in classical statistical physics discussed in Section \ref{microcan}. This will lead us to the notion of the Gibbs state.

\subsection{Gibbs state from a maximally entangled state}\label{Gibbs}

Consider a many-body quantum system with Hamiltonian $H$ and Hilbert space $\mathcal{H}$. We focus on the subspace of states whose energies lie within a narrow window $[E, E+\delta]$. Denote this quantum microcanonical shell by $\mathcal{H}_R$, and let $P_R$ be the orthogonal projector onto this subspace. The maximally mixed state on the shell, in accord with \eqref{eqprob},  is then given by
\begin{equation}
  \omega_R = \frac{P_R}{d_R},
\end{equation}
where $d_R = \Tr P_R$ is the dimension of $\mathcal{H}_R$. This state represents equal a priori weighting of all microstates compatible with the fixed energy window, directly paralleling the classical microcanonical ensemble.
\smallskip

We now decompose the global Hilbert space into a tensor product $\mathcal{H} = \mathcal{H}_S \otimes \mathcal{H}_E$, where $S$ denotes the subsystem of interest and $E$ the remaining degrees of freedom. The reduced density matrix of $S$ corresponding to the microcanonical shell is obtained by tracing out the environment:
\begin{equation}
  \omega_S = \Tr_E \, \rho_R .
\end{equation}
This reduced state is generally mixed even though $\rho_R$ is maximally mixed only within the restricted subspace $\mathcal{H}_R$. Importantly, $\rho_S$ already contains nontrivial structure reflecting the interplay between the subsystem Hamiltonian $H_S$ and the energy constraint imposed at the global level.
\smallskip

The canonical typicality (discussed above in Section \ref{typ}) states that for the overwhelming majority of pure states $\ket{\Psi} \in \mathcal{H}_R$ (with respect to the Haar measure), the reduced state of the subsystem is essentially indistinguishable from $\omega_S$:
\begin{equation}
  \rho_S(\Psi) = \Tr_E \ket{\Psi}\bra{\Psi} \approx \omega_S .
\end{equation}
Thus the reduced density matrix of a typical pure state drawn from the energy shell is extremely close to the reduced microcanonical state. Moreover, under broad conditions on the Hamiltonian and the energy width $\delta$, the latter is well-approximated by a Gibbs state of the subsystem:
\begin{equation}
  \omega_S \approx \frac{e^{-\beta H_S}}{Z_S(\beta)} ,
\end{equation}
for a suitable inverse temperature $\beta$ determined by the global energy. This establishes a purely quantum-mechanical mechanism through which statistical ensembles emerge from typical properties of pure states.
\smallskip

Indeed, maximizing $S(\rho_S)$ subject to the conditions $\Tr(\rho_S)=1$ and $\Tr(\rho_S H_S)=\langle H_S\rangle$ implemented via Lagrange multipliers gives
\begin{equation}
\rho_S^\star = \arg\max_{\rho_S}\left\{ -\Tr \rho_S \ln\rho_S - \lambda_0(\Tr\rho_S-1) - \beta(\Tr\rho_S H_S-\varepsilon)\right\}
= \frac{e^{-\beta H_S}}{Z_S(\beta)} .
\end{equation}
Thermodynamic potentials follow:
\begin{equation}\label{therpot}
F_S(\beta) \equiv -\frac{1}{\beta}\ln Z_S(\beta),\ \ 
\langle H_S\rangle_\beta = -\pdv{\ln Z_S}{\beta},\ \ 
S_{\mathrm{th}}(\beta) = \beta\langle H_S\rangle_\beta + \ln Z_S(\beta).
\end{equation}

\subsection{Entanglement entropy and thermodynamic entropy}

The entanglement between $S$ and $E$ in a typical pure state leads naturally to an entropy that coincides with the conventional thermodynamic entropy of the subsystem. 
Once a subsystem is described by a canonical (Gibbs) state, its entropy obeys the standard thermodynamic relations. 
For a system with Hamiltonian \(H\) at temperature \(T=1/\beta\), the canonical density matrix is
\[
\rho = \frac{e^{-\beta H}}{Z}, 
\qquad 
Z(\beta)=\Tr\, e^{-\beta H},
\]
where \(Z\) is the partition function. 
The Helmholtz free energy is defined as
\[
F(T) = -T \ln Z ,
\]
and the thermodynamic entropy follows from
\begin{equation}
S = -\left(\frac{\partial F}{\partial T}\right)_{V,N}.
\label{eq:canonicalEntropyThermo}
\end{equation}
Using standard identities,
\[
\langle H\rangle = -\frac{\partial}{\partial\beta}\ln Z ,
\]
one finds the equivalent familiar form
\begin{equation}
S = \beta\bigl(\langle H\rangle - F\bigr)
= \beta \langle H\rangle + \ln Z .
\label{eq:canonicalEntropy}
\end{equation}
\smallskip
Importantly, this thermodynamic entropy coincides exactly with the von Neumann entropy of the Gibbs state given by \eqref{therpot},
\begin{equation}
  S(\rho_S) = -\Tr\!\left( \rho_S \ln \rho_S \right)
  \simeq \beta \langle H_S \rangle + \ln Z_S ,
\end{equation}
establishing the equivalence between thermodynamic and quantum information  notions of entropy in the canonical ensemble. As an explicit illustration, let me mention the Schwinger pair production by a pulse of electric field where the von Neumann entropy of entanglement between the electrons and positrons was shown to be exactly equal to the Gibbs entropy of the produced system \cite{Florio:2021xvj}.
\smallskip

All of this implies that once entanglement with the environment becomes maximal, all familiar thermodynamic relations emerge automatically from quantum mechanics, without additional probabilistic assumptions. 
Therefore, thermodynamic entropy can be understood as entanglement entropy: the information lost about the subsystem when only local degrees of freedom (within the subsystem) are accessible. This provides a fully quantum foundation for the statistical behavior of macroscopic systems, linking pure-state quantum mechanics to ensemble thermodynamics without additional probabilistic postulates.

\subsection{Emergent irreversibility from reduced dynamics}

While global quantum evolution is exactly unitary and therefore reversible, the effective dynamics of a subsystem is not. Tracing over the environment leads to a reduced state $\rho_S(t)$ that generically evolves toward a stationary Gibbs state,
\begin{equation}
  \rho_S(t) \longrightarrow \omega_S(\beta)
  = \frac{e^{-\beta H_S}}{Z_S(\beta)},
\end{equation}
despite the absence of fundamental dissipation. The apparent irreversibility arises from the growth of entanglement between the subsystem and its environment: information about the initial state of $S$ becomes delocalized across the full system and cannot be recovered by local operations. Thus, thermodynamic irreversibility is an emergent phenomenon arising from restricted access to global degrees of freedom, rather than a modification of microscopic laws.

\section{Maximal Entanglement and Decoherence in Fock Space}

In this section we discuss the notion of phase space distribution in quantum theory, and describe the transformation from the usual
configuration-space (coordinate -- momentum) representation to the
occupation number -- phase representation. We also 
show how to compute reduced density matrices in both languages.  In
particular, we will see that tracing over the phase variable yields a
reduced density matrix that is diagonal in the occupation number. This result is of key importance in our quantum-informatic interpretation of the parton model \cite{Kharzeev:2017qzs,Kharzeev:2021nzh}. 

\subsection{Entanglement in Fock space}

Traditional statistical mechanics is formulated in terms of a classical phase space spanned by $N$ canonical coordinates and $N$ conjugate momenta, $\{q_i,p_i\}$, with macroscopic states defined by probability distributions over this $2N$-dimensional space. Thermodynamic ensembles are constructed by averaging over thin energy shells, and ergodicity and equilibration are ultimately tied to the assumed structure and continuity of phase space. While this framework has proven extraordinarily successful in classical systems, its foundations become conceptually problematic when carried over to quantum mechanics.
\smallskip

As emphasized by von Neumann \cite{vonNeumann1929}, the classical notion of phase space has no precise counterpart in quantum theory. Canonical coordinates and momenta do not commute, $[q_i,p_j]=i\hbar\delta_{ij}$, precluding the simultaneous specification of sharp trajectories and rendering the idea of a continuous phase-space distribution fundamentally ill-defined. Instead, quantum mechanics is naturally formulated in terms of operators acting on Hilbert space, and statistical descriptions must be expressed in terms of density matrices rather than phase-space probability measures. Quasi-probability distributions such as the Wigner function can be introduced, but they do not reinstate a classical phase-space picture and may take negative values, underscoring their auxiliary rather than fundamental status.

The viewpoint shift from classical phase space to the density matrix allows to use the notion of typicality discussed in Section \ref{rand}. Once the relevant description is a density matrix on a (possibly constrained) Hilbert space, concentration of measure implies that for overwhelmingly many pure states in a high-dimensional subspace (e.g.\ an energy shell), the reduced state of a small subsystem is close to a universal maximally entangled form. Indeed, Page's theorem (see Section \ref{page}) provides the sharpest quantitative expression of this idea: for a Haar-typical pure state on $\mathcal H_A\otimes\mathcal H_B$ with $\dim\mathcal H_B\gg \dim\mathcal H_A$, the reduced density matrix $\rho_A=\Tr_B|\psi\rangle\langle\psi|$ is, with high probability, nearly maximally mixed and its entanglement entropy is nearly maximal. In other words, thermality is typical in quantum mechanics precisely because generic pure states are highly entangled.

This perspective becomes particularly compelling in high-energy physics. The observables of interest in relativistic scattering are momentum distributions of partons, and the parton model is formulated not in phase space but in Fock space, where states are labeled by occupation numbers of quanta with given momenta, spins, and internal quantum numbers. In this setting, the relevant statistical object is the density matrix reduced to the occupation-number basis, obtained by tracing over unobserved degrees of freedom -- the relative phases. When the accessible sector is small compared to the unobserved complement -- as is the case at very high energies where the available Fock space grows rapidly -- typicality arguments suggest that the reduced state should approach the maximal entanglement limit. Understanding the structure of such reduced density matrices, and the conditions under which they become approximately diagonal and effectively thermal in the occupation-number representation, is therefore essential for a quantum-statistical description of high-energy interactions.
\medskip


Let us begin by considering a single bosonic degree of freedom with canonical operators
\(\hat x\) and \(\hat p\),
\begin{equation}
  [\hat x,\hat p] = i\hbar .
\end{equation}
In the coordinate representation,
\(|x\rangle\) are eigenstates of \(\hat x\),
\(\hat x |x\rangle = x |x\rangle\), and a pure state \(|\Psi\rangle\) is
represented by the wave function
\begin{equation}
  \Psi(x) \equiv \langle x|\Psi\rangle,
  \qquad
  \hat p = -i\hbar \frac{\partial}{\partial x} .
\end{equation}

To discuss reduced density matrices, we introduce a system \(S\) and an
environment \(E\), with (a set of) coordinates \(x\) and \(y\), respectively.  A
pure state \(|\Psi\rangle\) of the combined system is described in the
coordinate representation by
\begin{equation}
  \Psi(x,y) \equiv \langle x,y|\Psi\rangle ,
\end{equation}
and the corresponding density matrix is
\begin{equation}
  \hat\rho = |\Psi\rangle\langle\Psi| .
\end{equation}
Its matrix elements in the coordinate basis are
\begin{equation}
  \rho(x,y; x',y')
  \equiv \langle x,y|\hat\rho|x',y'\rangle
  = \Psi(x,y)\,\Psi^*(x',y') .
\end{equation}

The reduced density matrix of the system \(S\) is obtained by tracing
over the environmental coordinate \(y\),
\begin{equation}
  \hat\rho_S = \mathrm{Tr}_E\, \hat\rho .
\end{equation}
In the coordinate representation this reads
\begin{equation}
  \rho_S(x,x')
  \equiv \langle x|\hat\rho_S|x'\rangle
  = \int dy\, \langle x,y|\hat\rho|x',y\rangle
  = \int dy\, \rho(x,y; x',y) .
  \label{eq:rhoS_coordinate}
\end{equation}
Equation~\eqref{eq:rhoS_coordinate} is the standard expression for
partial tracing in the coordinate representation.
\medskip

We now introduce the harmonic oscillator creation and annihilation operators \(\hat a\) and
\(\hat a^\dagger\) via
\begin{equation}
  \hat a = \sqrt{\frac{m\omega}{2\hbar}}\,\hat x
          + \frac{i}{\sqrt{2m\hbar\omega}}\,\hat p,
  \qquad
  \hat a^\dagger = \sqrt{\frac{m\omega}{2\hbar}}\,\hat x
          - \frac{i}{\sqrt{2m\hbar\omega}}\,\hat p,
\end{equation}
where $m$ is the mass of the particle and $\omega$ is the oscillator frequency. 

\noindent The number operator is 
\begin{equation}
  \hat n = \hat a^\dagger \hat a .
\end{equation}
The eigenstates \(|n\rangle\) of \(\hat n\),
\begin{equation}
  \hat n |n\rangle = n |n\rangle, \qquad n=0,1,2,\dots ,
\end{equation}
provide the occupation--number basis.  The coordinate and number
representations are related by
\begin{equation}
  \phi_n(x) \equiv \langle x|n\rangle ,
\end{equation}
where \(\phi_n(x)\) are the usual harmonic-oscillator eigenfunctions.
For the combined system \(S+E\), we can use the tensor-product number
basis \(|n_S,n_E\rangle = |n_S\rangle\otimes|n_E\rangle\).

Given the reduced density matrix \(\hat\rho_S\) in the coordinate
representation, its matrix elements in the number basis are
\begin{equation}
  \rho_S(n,n')
  \equiv \langle n|\hat\rho_S|n'\rangle
  = \int dx\,dx'\, \phi_n^*(x)\,\rho_S(x,x')\,\phi_{n'}(x') .
  \label{eq:rhoS_number_from_coordinate}
\end{equation}
Equations~\eqref{eq:rhoS_coordinate} and
\eqref{eq:rhoS_number_from_coordinate} together describe the full
transformation from the coordinate representation of the total state to
the reduced density matrix in the occupation--number representation.

\subsection{Phase states and number--phase Fourier transform}

The conjugate variable to the occupation number is the phase \(\phi\).
Because \(\hat n\) has a discrete spectrum bounded from below, a
self-adjoint phase operator does not exist on the full Fock space.
However, one can construct phase states \(|\phi\rangle\) that are
Fourier dual to number states.  In the approach proposed by Pegg and Barnett \cite{Pegg:1989zz} one
defines, in a truncated Fock space of dimension \(s+1\),
\begin{equation}
  |\phi_m^{(s)}\rangle
  = \frac{1}{\sqrt{s+1}}\sum_{n=0}^{s} e^{i n\phi_m} |n\rangle,
  \qquad \phi_m = \phi_0 + \frac{2\pi m}{s+1},\quad
  m=0,\dots,s .
\end{equation}
In the limit \(s\to\infty\), this construction yields continuous phase
states \(|\phi\rangle\) with
\begin{equation}
  \langle \phi|n\rangle = \frac{1}{\sqrt{2\pi}}\, e^{-i n\phi},
  \qquad \phi\in[0,2\pi),
\end{equation}
and the (formal) completeness relation
\begin{equation}
  \int_0^{2\pi} d\phi\, |\phi\rangle\langle\phi|
  = \mathbbm{1}.
\end{equation}
These relations are the exact analog of the plane-wave decomposition in
the coordinate--momentum case.
\smallskip

For a state \(|\psi\rangle\) of a single mode, we have
a number representation:
\begin{equation}
  \psi(n) \equiv \langle n|\psi\rangle,
\end{equation}

\noindent
and a phase representation:
\begin{equation}
  \psi(\phi) \equiv \langle \phi|\psi\rangle
  = \frac{1}{\sqrt{2\pi}}\sum_{n=0}^{\infty} \psi(n)\,e^{i n\phi}.
  \label{eq:number_to_phase}
\end{equation}
Conversely,
\begin{equation}
  \psi(n)
  = \frac{1}{\sqrt{2\pi}}
    \int_0^{2\pi} d\phi\, e^{-i n\phi}\,\psi(\phi).
  \label{eq:phase_to_number}
\end{equation}
Equations~\eqref{eq:number_to_phase} and \eqref{eq:phase_to_number} show
that number and phase are related by a Fourier transform on the circle.

\subsection{Tracing over the phase and diagonality in occupation number}\label{fock-dec}

We now show how tracing over the phase variable yields a reduced density
matrix that is diagonal in the occupation number.  Physically, this
corresponds to losing all information about the conjugate phase degree
of freedom, or equivalently, averaging over global \(U(1)\) rotations
generated by \(\hat n\) with a corresponding Haar measure.
\smallskip

Consider an arbitrary density matrix \(\hat\rho\) expressed in the
number basis:
\begin{equation}
  \hat\rho = \sum_{n,n'=0}^{\infty}
             \rho_{n n'}\,|n\rangle\langle n'| .
\end{equation}
We define the phase-averaged (or phase-reduced) density matrix
\(\hat\rho_{\text{red}}\) by integrating over the phase with the \(U(1)\) Haar measure $d\phi/2\pi$,
\begin{equation}
  \hat\rho_{\text{red}}
  = \frac{1}{2\pi}\int_0^{2\pi} d\phi\,
    e^{i\phi\hat n}\,\hat\rho\,e^{-i\phi\hat n}.
  \label{eq:phase_average_def}
\end{equation}
This operation can be viewed as a trace over the unobserved phase
variable, or as a projection onto the \(U(1)\)-invariant part of
\(\hat\rho\).
\smallskip

To see its effect in the number basis, note that
\begin{equation}
  e^{i\phi\hat n}|n\rangle = e^{i\phi n}|n\rangle ,
  \qquad
  \langle n'|e^{-i\phi\hat n} = e^{-i\phi n'}\langle n'|.
\end{equation}
Therefore
\begin{equation}
  e^{i\phi\hat n}\,|n\rangle\langle n'|\,e^{-i\phi\hat n}
  = e^{i\phi(n-n')}\,|n\rangle\langle n'|.
\end{equation}
Substituting into \eqref{eq:phase_average_def} yields
\begin{equation}
  \hat\rho_{\text{red}}
  = \sum_{n,n'} \rho_{n n'}\,
    \left[ \frac{1}{2\pi}
    \int_0^{2\pi} d\phi\; e^{i\phi(n-n')} \right]
    |n\rangle\langle n'| .
\end{equation}
Using
\begin{equation}
  \frac{1}{2\pi}\int_0^{2\pi} d\phi\; e^{i\phi(n-n')}
  = \delta_{n n'},
\end{equation}
we obtain
\begin{equation}
  \hat\rho_{\text{red}}
  = \sum_{n} \rho_{n n}\,|n\rangle\langle n|.
  \label{eq:rho_red_diagonal}
\end{equation}
Thus the phase-averaged (phase-traced) density matrix is exactly
diagonal in the occupation number.

Equivalently, one can express this as a partial trace in the
number--phase representation:
\begin{equation}
  \rho_{\text{red}}(n,n')
  \equiv \langle n|\hat\rho_{\text{red}}|n'\rangle
  = \int_0^{2\pi} d\phi\;
    \langle n,\phi|\hat\rho|n',\phi\rangle,
\end{equation}
where the ``trace over phase'' is represented by the integral over
\(\phi\).  The Fourier relations between number and phase guarantee that
only the diagonal terms in \(n\) survive, as in
\eqref{eq:rho_red_diagonal}.

\section{Entanglement as the Origin of the Parton Model}\label{opm}

High-energy interactions, following the pioneering work of Feynman \cite{Feynman:1969ej}, Bjorken \cite{Bjorken:1968dy}, and Gribov \cite{Gribov1969}, are traditionally formulated in terms of the parton model -- a framework that is explicitly probabilistic in nature. This raises a fundamental question: how does a probabilistic parton description emerge from the underlying unitary quantum dynamics of QCD? 
\smallskip

This is nothing but the Boltzmann's original question in a new context: why should a deterministic microscopic theory (QCD) give rise to a probabilistic description (parton model)? In classical statistical mechanics, Boltzmann answered this by appealing to typicality in phase space -- an assumption of equal probabilities for all microstates compatible with macroscopic constraints. In high-energy physics, however, the relevant description is not formulated in classical phase space, but in Fock space on the light cone. 
\smallskip

The central idea explored here, following Refs.~\cite{Kharzeev:2017qzs,Kharzeev:2021nzh}, is that the probabilistic parton model arises in precisely the same way as Boltzmann's ensembles: as a typical reduced description of an underlying pure quantum state. The role of phase-space averaging is now played by tracing over unobservable light-cone time \cite{Kharzeev:2021nzh}, and the emergence of probabilities is driven not by classical ignorance, but by quantum entanglement. I will now make this statement explicit by showing how the light-cone density matrix becomes diagonal in the occupation-number basis once light-cone time is traced out, and how the resulting reduced density matrix is nothing but the familiar probabilistic parton model.

\subsection{Parton model from high-energy decoherence}\label{decoh}

High-energy scattering processes are most naturally formulated in light-cone coordinates introduced by Dirac \cite{Dirac:1949cp},
\[
x^\pm = \frac{1}{\sqrt{2}}(t \pm z), 
\qquad 
p^\pm = \frac{1}{\sqrt{2}}(E \pm p_z),
\]
where the light-cone time $x^+$ plays the role of the evolution parameter and $p^+$ is kinematically constrained to be positive. A fast-moving hadron is therefore described by a quantum state defined on a light-like hypersurface $x^+ = \mathrm{const}$, which admits a Fock-space expansion in terms of partonic degrees of freedom (see \cite{Brodsky:1997de} for a review),
\[
\ket{\Psi} = \sum_{n} 
\int [dx_i\, d^2 k_{\perp i}]\,
\psi_n(x_i, k_{\perp i})\,
\ket{n; x_i, k_{\perp i}},
\]
where $x_i = p_i^+/P^+$ are longitudinal momentum fractions, and $k_{\perp i}$ are the transverse momenta of partons. These basis states are eigenstates of the parton number operators and of the light-cone Hamiltonian $P^-$.
\smallskip

The full quantum state is pure, with density matrix
\[
\rho = \ket{\Psi}\bra{\Psi},
\]
and therefore contains coherent superpositions between different Fock sectors and momentum configurations. However, physical observables measured in high-energy experiments -- such as inclusive cross sections in deep inelastic scattering -- are insensitive to the light-cone time $x^+$. Indeed, the light-cone time cannot be measured during a high-energy interaction. In the usual $(t,z)$ space, this is due to Lorentz contraction in the fast-moving frame. 

Indeed, the parton model is formulated in a frame where the projectile carries very large longitudinal momentum $p^+$. In this limit, the light--cone energy of its internal excitations,
\begin{equation}
p^- = \frac{m^2 + p_\perp^2}{2p^+},
\end{equation}
is parametrically small. The characteristic time scale governing the internal quantum evolution of the fast projectile is therefore
\begin{equation}
\Delta x^+ \sim \frac{1}{\Delta p^-} \;\propto\; p^+ ,
\end{equation}
which becomes much larger than the duration of the interaction. Physically, Lorentz time dilation freezes the internal dynamics of the projectile during the scattering, rendering any dependence on $x^+$ inaccessible to measurement.

From the viewpoint of the probe, hard interactions are localized in light--cone position $x^-$ over a short interval,
\begin{equation}
\Delta x^- \sim \frac{1}{q^+},
\end{equation}
with no experimental sensitivity to the conjugate variable $x^+$ 
(in deep-inelastic scattering, $q^+$ 
 is the light-cone momentum of the exchanged virtual photon). As a result, observables are insensitive to the light--cone time, and phase factors of the form $\exp(-i p^- x^+)$ associated with different Fock components are averaged over and become unobservable.
\smallskip

The inobservability of $x^+$ means that we have to trace over it, or equivalently, over the phases generated by the light-cone Hamiltonian, in the density matrix:
\[
\rho_{\mathrm{red}} = \mathrm{Tr}_{x^+}\, \rho .
\]
This trace suppresses off-diagonal matrix elements through rapid phase oscillations of the form
\[
\exp\!\left[-i (P^-_n - P^-_m) x^+ \right],
\]
which average to zero whenever the participating states carry different light-cone energies. 
This is exactly the Fock-space decoherence discussed above in Section \ref{fock-dec}.
\smallskip

As a result, interference between Fock states carrying different $p^-$ is suppressed, and the reduced density matrix relevant for high--energy observables becomes diagonal in the occupation--number basis \cite{Kharzeev:2021nzh},
\be\label{partm}
\rho_{\mathrm{red}} =
\sum_n P_n\, \ket{n}\bra{n},
\qquad
P_n = \int [dx_i\, d^2 k_{\perp i}]\,
|\psi_n(x_i, k_{\perp i})|^2 .
\ee

This diagonal form admits a natural interpretation in terms of entanglement. Tracing over the unobservable light-cone time generates entanglement entropy \cite{Kharzeev:2017qzs,Kharzeev:2021nzh},
\[
S_{\mathrm{LC}} = -\mathrm{Tr}\left(\rho_{\mathrm{red}} \ln \rho_{\mathrm{red}}\right) > 0,
\]
even though the underlying hadronic state is pure. The entropy quantifies the loss of phase information associated with inaccessible degrees of freedom (phases) and is directly tied to the incoherent nature of the observed partonic ensemble. The emergence of a probabilistic description on the light cone is thus a manifestation of quantum decoherence driven by entanglement.
\smallskip

This perspective resonates closely with von Neumann's quantum ergodic theorem, which states that for a generic pure state evolving under a complex Hamiltonian, the reduced density matrix of a small subsystem is, for most times, indistinguishable from a microcanonical ensemble. On the light cone, the role of time averaging is played by tracing over light-cone time $x^+$, while the ``subsystem'' corresponds to the momentum-space degrees of freedom probed in high-energy scattering. The diagonal density matrix $\rho_{\mathrm{red}}$ is thus a concrete realization of quantum typicality in Fock space.
\smallskip

Finally, this construction provides a direct bridge to phenomenology, since the matrix elements of $\rho_{\mathrm{red}}$ reproduce the probabilistic parton model underlying the description of deep inelastic scattering. Structure functions are obtained as expectation values of appropriate number operators weighted by $P_n$, with no sensitivity to relative phases between Fock components. 
\smallskip

To see explicitly how the reduced density matrix reproduces the standard parton-model recipe, consider an observable $\widehat{\mathcal O}$ that is insensitive to light-cone time and is diagonal in the partonic occupation basis (as is the case for inclusive observables built from number operators). The expectation value of  $\widehat{\mathcal O}$ over the density matrix \eqref{partm} is
\[
\langle \widehat{\mathcal O}\rangle
=\Tr\!\left(\rho_{\rm red}\,\widehat{\mathcal O}\right)
=\sum_{n} P_n\,\bra{n}\widehat{\mathcal O}\ket{n}.
\]
For a one-body inclusive operator probing partons of species $a$, one may take
\[
\widehat{\mathcal O}_a(\phi)\;=\;\sum_{i\in a}\,\phi(x_i,k_{\perp i}),
\]
where $\phi$ is the partonic weight appropriate to the measured quantity (e.g.\ a projector onto a bin in $x$ and $k_\perp$). Then
\[
\begin{aligned}
\langle \widehat{\mathcal O}_a(\phi)\rangle
&=\sum_{n}\int [dx_i\,d^2k_{\perp i}]\,|\psi_n|^2\,
\sum_{i\in a}\phi(x_i,k_{\perp i}) \\
&=\int_0^1\!dx\int\!d^2k_\perp\;\phi(x,k_\perp)\,f_a(x,k_\perp),
\end{aligned}
\]
where the (unintegrated) parton distribution emerges as
\[
f_a(x,k_\perp)\;=\;\sum_{n}\int [dx_i\,d^2k_{\perp i}]\,|\psi_n(x_i,k_{\perp i})|^2
\sum_{i\in a}\delta(x-x_i)\,\delta^{(2)}(k_\perp-k_{\perp i}) .
\]
Thus the quantum expectation value computed with $\rho_{\rm red}$ reduces to the standard parton-model rule: inclusive observables are obtained by averaging over incoherent partonic configurations with probabilities $|\psi_n|^2$, or equivalently by integrating the parton distribution $f_a$ with the weight $\phi$.
\smallskip

The familiar incoherent picture of partons carrying definite momentum fractions therefore emerges naturally as a reduced description of an underlying coherent quantum state. From this viewpoint, the parton model is not an assumption, but the consequence of tracing over unobserved phases in a highly entangled quantum system.  In other words, the parton model is what remains when quantum coherence is lost.

\subsection{When parton model breaks down: spin puzzles and zero modes}

A central reason why the parton model provides an accurate description of high--energy scattering is that, due to extreme Lorentz time dilation, the relative phases of different Fock space components of a fast hadron become unobservable in any finite--time measurement. As a result, interference effects between Fock states are effectively washed out, and the hadronic state is operationally described by a diagonal density matrix in the occupation--number basis. In this limit, probabilities rather than amplitudes determine observables, and the parton model emerges as a valid and efficient approximation.

At low energies, the situation is qualitatively different. In elastic and quasi--elastic scattering at low momentum transfer, quantum coherence is preserved and the phases of scattering amplitudes are directly observable. Indeed, the low--energy $S$--matrix is conventionally parameterized in terms of phase shifts, reflecting the sensitivity of observables to relative phases between partial waves. In this regime, phase information is physically meaningful and cannot be neglected.

The observability of phases has direct implications for the interpretation of occupation numbers. Number and phase are canonically conjugate variables, obeying an uncertainty relation of the form
$$
  \Delta n \, \Delta \phi \;\gtrsim\; 1 ,
$$
where $n$ denotes the occupation number of a given mode and $\phi$ its conjugate phase. When phases are sharply defined, as in low--energy coherent scattering, the uncertainty in occupation numbers necessarily becomes large. Consequently, the notion of well--defined parton densities loses its meaning, and the probabilistic interpretation underlying the parton model breaks down. The validity of the parton picture at high energies and its failure at low energies can thus be viewed as complementary manifestations of the number--phase uncertainty relation.
\bigskip

While unpolarized inclusive observables at high energies are largely insensitive to quantum phases and admit a probabilistic parton interpretation, this is no longer true for a broad class of polarization--dependent measurements. Spin and polarization observables are intrinsically interference effects: they arise from correlations between different helicity or transverse--spin amplitudes and therefore depend explicitly on their relative phases. As a result, even in the high--energy limit where Lorentz time dilation suppresses phase sensitivity in unpolarized cross sections, polarization observables can render these phases experimentally accessible.

When such phase information becomes observable, the reduced density matrix describing the relevant degrees of freedom is no longer diagonal in the occupation--number basis. Off--diagonal coherence terms survive and contribute directly to measured asymmetries, such as single--spin and double--spin observables. Through the number--phase uncertainty relation, the partial restoration of phase coherence implies an increased uncertainty in occupation numbers, undermining the probabilistic interpretation of parton densities. The breakdown of the naive parton model in polarization phenomena thus reflects a fundamental quantum limitation rather than a failure of perturbative dynamics, and highlights the necessity of amplitude--level or density--matrix--based descriptions for spin--dependent high--energy processes.

This sensitivity of polarization observables to quantum phases, and the resulting breakdown of a purely probabilistic description in terms of parton occupation numbers, provides a likely underlying mechanism for the numerous ``spin puzzles'' encountered within the parton model framework.
\bigskip

Another reason for the possible breakdown of the parton model is 
the presence of degeneracies or near--degeneracies in the light--cone energy spectrum. If two components satisfy
\begin{equation}
\Delta p^- \simeq 0 ,
\end{equation}
their relative phase remains coherent over parametrically long light--cone times,
\begin{equation}
\Delta x^+ \sim \frac{1}{\Delta p^-} \;\to\; \infty ,
\end{equation}
and interference terms survive the averaging over $x^+$. In this situation, the reduced density matrix is no longer diagonal in the parton basis, and the probabilistic parton model breaks down. The obstruction is therefore not dynamical but kinematical: phase coherence persists because the light--cone energies fail to separate.

This observation provides a natural perspective on the long--standing difficulty of incorporating the nonperturbative QCD vacuum into the parton model. The QCD vacuum phenomena are associated with degrees of freedom whose light--cone energies are zero or ill--defined. In light--cone quantization, these appear as zero modes, characterized by vanishing longitudinal momentum $p^+=0$, for which the expression for $p^-$ becomes singular or ambiguous. Such modes do not decouple from the dynamics and cannot be treated as incoherent partons with well--defined occupation probabilities.

Zero modes encode long--range correlations, topological structure, and collective phenomena such as chiral symmetry breaking and confinement. Their presence implies an extensive degeneracy in light--cone energies, preventing the loss of phase information that underlies the parton picture. This places the nonperturbative QCD vacuum outside the domain of validity of the parton model and explains why full density--matrix formulations are required.

\section{The Maximal Entanglement Limit of High-Energy QCD}


In the large $N_c$ limit, QCD evolution at small Bjorken \(x\) as described by BFKL equation \cite{Fadin:1975cb,Balitsky:1978ic} admits a particularly transparent formulation in terms of the color dipole model, as originally proposed by Mueller \cite{Mueller:1993rr}. In this framework, the light--cone wave function of a fast hadron or virtual photon is represented as an ensemble of color dipoles in transverse coordinate space, whose evolution in rapidity \(Y=\ln(1/x)\) proceeds through successive dipole splittings. 
\smallskip

Assume a branching process in which dipole evolution in rapidity is Markovian and depends on the state only through the current multiplicity \(n\). It is convenient to analyze this evolution via the multiplicity generating function \cite{Mueller:1994gb}
\[
Z(Y,u)\equiv \sum_{n=1}^\infty P_n(Y)\,u^n, 
\]
where \(P_n(Y)\) denotes the probability to find \(n\) dipoles at rapidity \(Y\). It obeys the boundary conditions
\[
Z(Y=0,u)=u , \ \ Z(Y,u=1)=1.
\]
The first of these relations is a straightforward consequence of the initial condition $P_1(Y=0)=1$, $P_{n>1}(Y=0)=0$ which assumes that the evolution begins  with a single dipole (a pure state). The second one is just the normalization of the multiplicity distribution $\sum_n P_n(Y) =1$ that holds at all $Y$.

For a wide class of such splitting processes the multiplicity generating function obeys a closed equation of the logistic type
\be\label{logis}
\partial_Y Z(Y,u)=\Delta\bigl(Z^2(Y,u)-Z(Y,u)\bigr),
\qquad Z(0,u)=u,
\ee
where \(\lambda \sim \bar\alpha_s\) sets the growth rate; $\bar\alpha_s \equiv \frac{\alpha_s N_c}{\pi}$ is the rescaled QCD coupling. 
This equation can be solved exactly:
\[
Z(Y,u)=\frac{u}{u+(1-u)e^{\lambda Y}}.
\]
Introducing the mean multiplicity \(a(Y)\equiv \langle n\rangle = e^{\lambda Y}\), and re-writing\\ $u+(1-u)a = a-(a-1)u$, this solution can be written as
\[
Z(Y,u)=\frac{u}{a-(a-1)u}
= \frac{1}{a}\,\frac{u}{1-\Bigl(1-\frac{1}{a}\Bigr)u}.
\]
Expanding in powers of \(u\) gives \cite{Levin:2003nc}
\[
Z(Y,u)=\sum_{n=1}^{\infty}\left[\frac{1}{a}\left(1-\frac{1}{a}\right)^{n-1}\right]u^n,
\]
hence the exact multiplicity distribution is geometric \cite{Kharzeev:2017qzs}:
\be\label{geomd}
P_n(Y)=\frac{1}{a(Y)}\left(1-\frac{1}{a(Y)}\right)^{n-1},
\qquad n=1,2,\ldots,
\qquad \langle n\rangle=a(Y).
\ee
\smallskip

Let us show that at high multiplicity, in the scaling regime \(a\to\infty\), \(n\to\infty\) with \(n/a\) fixed, this distribution becomes Boltzmann-like maximum entropy distribution. 
Using
\[
\ln\!\left(1-\frac{1}{a}\right)=-\frac{1}{a}-\frac{1}{2a^2}-\frac{1}{3a^3}-\cdots,
\]
one finds
\[
\left(1-\frac{1}{a}\right)^{n-1}
=\exp\!\left[-\frac{n-1}{a}-\frac{n-1}{2a^2}+\mathcal{O}\!\left(\frac{n}{a^3}\right)\right].
\]
In the scaling regime \(a\to\infty\), \(n\to\infty\) with \(n/a\) fixed,
\be\label{expdist}
P_n(Y)=\frac{1}{a}\left(1-\frac{1}{a}\right)^{n-1}
\;\simeq\;\frac{1}{a}\,e^{-n/a}
\left[1+\mathcal{O}\!\left(\frac{n}{a^2}\right)\right],
\ee
so the discrete geometric distribution becomes the continuous maximum-entropy exponential distribution. Note that this exponential approximation holds only at large multiplicity {\it and} large $n$. The effective temperature of this Boltzmann-like distribution grows with rapidity, $T_{eff} \sim \exp(\lambda Y)$ (or equivalently as a power of energy).
\smallskip

The emergence of a continuous exponential multiplicity distribution at large rapidity can be understood as a particular realization of Koba--Nielsen--Olesen (KNO) scaling \cite{Koba:1972ng}. According to KNO scaling (proposed in 1970 by Polyakov  \cite{Polyakov:1970lyy}), the multiplicity distribution depends on energy (or rapidity) only through the mean multiplicity \(\langle n\rangle\), such that
\[
P_n(Y)=\frac{1}{\langle n(Y)\rangle}\,
\psi\!\left(\frac{n}{\langle n(Y)\rangle}\right),
\]
with a universal scaling function \(\psi(z)\). The guiding principle that led Polyakov to this result \cite{Polyakov:1970lyy} was the asymptotic scaling invariance of strong interactions at short distances; note that this paper was written before the asymptotic freedom \cite{Gross:1973id,Politzer:1973fx} became a universally accepted concept. 

Polyakov  also predicted \cite{Polyakov:1970lyy} a power growth of mean multiplicity with energy, or, equivalently, an exponential growth in rapidity discussed above. This is not a coincidence, because the dipole evolution equation considered above is a realization of renormalization group flow in a scale-invariant field theory -- indeed, the splitting kernel of (\ref{logis}) is a dimensionless constant and is trivially scale-invariant. The emergence of KNO scaling in the dipole formulation has been investigated in \cite{Liu:2022bru}.
The consequences of asymptotic freedom for multiplicity distributions have been recently discussed by Dokshitzer and Webber  \cite{Dokshitzer:2025owq,Dokshitzer:2025fky}. \smallskip

The exponential distribution (\ref{expdist}) obtained above from the dipole cascade,
\be\label{expdistr}
P_n(Y)\simeq \frac{1}{\langle n(Y)\rangle}\,
\exp\!\left(-\frac{n}{\langle n(Y)\rangle}\right),
\ee
corresponds to the specific choice \(\psi(z)=e^{-z}\).
This form represents the maximum-entropy scaling function compatible with KNO scaling and fixed \(\langle n\rangle\). In this sense, the approach to an exponential multiplicity distribution in small-\(x\) QCD signals not only the onset of KNO scaling, but its approach to the most entropic form.
\smallskip

It is useful to contrast the geometric law with two standard multiplicity models:
\smallskip

(i) Poisson distribution: For independent emissions at fixed rate one obtains
$$
P_n^{\rm Poisson}(a)=e^{-a}\ \frac{a^n}{n!},
\qquad
\mathrm{Var}(n)=a=\langle n\rangle,
$$
which is much more narrow than (\ref{expdistr}). Note that Poisson distribution is characteristic for the statistics of quanta in a classical field. The entropy in this case is much smaller for a given mean $\langle n\rangle$.  
\smallskip

Therefore, the classical field, Weizs\"acker--Williams limit is in contrast to the Maximal Entanglement Limit (MEL). Below in Section \ref{cgc} we discuss the ensuing relation between the MEL and the Color Glass Condensate approaches to high-energy interactions. 
\smallskip

(ii) Negative binomial (NB) distribution: A broad class of cascading and ``clustering'' mechanisms lead to
$$
P_n^{\rm NB}(k,\bar n)=
\frac{\Gamma(n+k)}{\Gamma(k)\,\Gamma(n+1)}
\left(\frac{\langle n\rangle}{\langle n\rangle+k}\right)^n
\left(\frac{k}{\bar n+k}\right)^k,
\qquad
\mathrm{Var}(n)=\bar n+\frac{\langle n\rangle^2}{k}.
$$
The NB law interpolates between Poisson and geometric:
$$
k\to\infty \ \Rightarrow\ P_n^{\rm NB}\to P_n^{\rm Poisson}(\bar n),
$$
and 
$$
k=1 \ \Rightarrow\ P_n^{\rm NB}(1,\langle n\rangle)
=\frac{1}{1+\langle n\rangle}\left(\frac{\langle n\rangle}{1+\langle n\rangle}\right)^n,
$$
which is the geometric distribution, up to the conventional support \(n=0,1,2,\ldots\) rather than \(n\ge 1\).
Thus, the geometric law corresponds to the maximally broad NB case \(k=1\), characteristic of strong fluctuations in a branching process, whereas deviations from this limit (effective \(k>1\)) quantify suppression of fluctuations, e.g.\ due to classical coherence. The emergence of NB distribution in the dipole cascade, and the corresponding quantum information measures, were recently discussed in \cite{Kutak:2025syp}.
\bigskip

The entanglement entropy should be identified with the Shannon entropy of the multiplicity distribution,
\begin{equation}
S(Y) = - \sum_{n} P_n(Y)\,\ln P_n(Y),
\end{equation}
since the reduced density matrix is diagonal in the occupation number basis (the parton number basis is the Schmidt basis\footnote{The Schmidt theorem states that any pure quantum state of a bipartite system can be written in a basis where the reduced density matrices of both subsystems are diagonal and share the same nonzero eigenvalues, which quantify their entanglement.}). 
For both the geometric distribution and its continuous exponential limit, the entropy is dominated at large rapidity $Y$ by
\begin{equation}
S(Y) \simeq \ln \langle n(Y) \rangle + \mathcal{O}(1) \simeq \lambda Y ,
\end{equation}
demonstrating that the entanglement entropy grows linearly with rapidity \cite{Kharzeev:2017qzs}. This linear growth reflects the exponential proliferation of dipoles and the accompanying increase in the number of entangled degrees of freedom generated by the branching process.
\smallskip

From the viewpoint of the Maximal Entanglement Limit (MEL), this result acquires a natural interpretation. The geometric distribution (with its exponential Boltzmann limit) is the maximum--entropy distribution for a positive integer variable at fixed mean multiplicity. This means that, at large rapidity, the dipole ensemble saturates the entropy allowed by its single constraint, the mean multiplicity.  
Therefore, the linear growth of entropy with rapidity is a direct manifestation of the system approaching the MEL: high--energy evolution drives the system toward a universal, maximally entangled state in which an exponentially (in rapidity) large number of dipole micro-states have approximately equal and exponentially small probabilities \cite{Kharzeev:2017qzs}.
\smallskip

It is important to point out that the linear growth of entanglement entropy with rapidity is not a peculiarity of dipole branching models, but a much more general consequence of approximate scale invariance at high energies. As shown in \cite{Gursoy:2023hge}, this behavior persists even in string--based descriptions of high--energy scattering, where the microscopic degrees of freedom are fundamentally different from those of perturbative partons or dipoles. In such models, the rapidity evolution maps naturally onto a scale--invariant dynamics along the string worldsheet, leading to an extensive growth of entanglement between longitudinally separated segments. As a result, the entanglement entropy takes the universal form
\begin{equation}
S(Y) \simeq c \, Y ,
\end{equation}
with the coefficient $c$ proportional to the effective central charge, but independent of the detailed microscopic realization. 

In the MEL, the high--energy behavior is thus fixed by the central charge of the corresponding conformal theory. Recently, using the Lipatov effective high--energy theory \cite{Lipatov:1993yb,Faddeev:1994zg}, this was used to compute the small $x$ dependence of the gluon structure function \cite{Grieninger:2025wxg} -- the central charge $c=1$ of the Lipatov theory \cite{Zhang:2021hra,Hao:2019cfu} was found to translate to $xG(x) \sim x^{-c/3} \sim x^{-1/3}$, in good agreement with phenomenology.
\smallskip

The linear dependence of entanglement entropy on rapidity admits a natural and unifying interpretation in terms of self--similar renormalization--group (RG) flow. In high--energy scattering, rapidity $Y$ plays the role of an RG ``time'': increasing $Y$ corresponds to integrating out quantum fluctuations with progressively shorter lifetimes. When the underlying dynamics is approximately scale invariant, this RG flow is self--similar, meaning that each infinitesimal step in rapidity generates statistically equivalent degrees of freedom.

Entanglement entropy is produced whenever degrees of freedom are integrated out. In a scale--invariant theory, the amount of information lost -- and hence the entanglement generated -- per RG step is constant. This immediately implies
\begin{equation}
\frac{dS}{dY} = \text{const},
\end{equation}
where the constant is fixed by the scaling data of the theory, rather than by microscopic details of the dynamics. Linear growth of entanglement entropy in rapidity \cite{Kharzeev:2017qzs} is therefore a direct and robust consequence of scale invariance.

This structure closely parallels well--known results in conformal field theory. In a $(1+1)$--dimensional conformal field theory (CFT), the entanglement entropy of an interval of length $L$ scales as \cite{Holzhey:1994we,Calabrese:2004eu}
\begin{equation}
S(\ell) = \frac{c}{3}\,\ln\left(\frac{L}{\varepsilon}\right),
\end{equation}
where $c$ is the central charge, which measures the effective number of (gapless) degrees of freedom, and $\varepsilon$ is a short--distance (UV) cutoff. In the high--energy context, rapidity $Y$ plays the role of a logarithmic scale variable, and the coefficient $\kappa$ governing entropy production per unit rapidity is the analogue of an effective central charge \cite{Kharzeev:2017qzs}. It quantifies how many independent degrees of freedom are generated and entangled at each step of the RG flow.
The linear growth of entropy with rapidity is thus the universal signature of a system approaching MEL under a self--similar RG flow. 

This connection can be made explicit by the following argument \cite{Kharzeev:2017qzs}. When viewed in the target rest frame, deep inelastic scattering at small Bjorken $x$ develops over the Ioffe coherence length \cite{Gribov:1965hf,Ioffe:1969kf}
$L \;\simeq\; 1/(m x)$, 
while the natural UV cutoff is set by the Compton wavelength of the proton of mass $m$,
$
\varepsilon \;\simeq\; 1/m.
$
Substituting these scales into the CFT expression yields
\begin{equation}
S = \frac{c}{3}\,\ln\!\left(\frac{1/mx}{1/m}\right)
= \frac{c}{3}\,\ln\!\left(\frac{1}{x}\right)
\;\equiv\; \kappa\, Y ,
\end{equation}
with $Y=\ln(1/x)$ the rapidity. 
\smallskip

This establishes a direct correspondence between the logarithmic entanglement growth in spatial scale for a CFT and the linear growth of entropy with rapidity in small--$x$ evolution. From the MEL perspective, this analogy highlights a common underlying mechanism: approximate scale invariance ensures that each logarithmic interval in length, or equivalently each unit of rapidity, contributes a fixed amount of entanglement. The coefficient $\kappa$, playing the role of an effective central charge, measures the number of active degrees of freedom generated and entangled per step in rapidity.

\subsection{From classical fields to maximal entanglement:\\ Color Glass Condensate vs MEL}\label{cgc}

At high energies, QED interactions are accurately described by the Weizs\"acker--Williams, or equivalent-photon, approximation. In simple terms, the boosted Coulomb field of a charge moving with a high velocity can be represented by a superposition of nearly on-shell photons that are emitted independently and thus possess the Poisson statistics -- at high enough occupation number, this is a classical electromagnetic field. This classical field limit contradicts the Maximal Entanglement Limit (MEL) advocated above for all high-energy interactions. Let us discuss this apparent contradiction, and the transition to MEL in QCD, in more detail. 
\smallskip


At moderately high energies, the Weizs\"acker--Williams regime holds, and observables can be expressed in terms of an effective quasi-real photon distribution, so the dynamics admits a quasi--classical probabilistic interpretation.
However, as originally discovered by Gribov and Lipatov \cite{Gribov:1972rt}, this description inevitably breaks down at sufficiently high energies, when the available longitudinal phase space becomes large enough to allow splittings of photons into  electron--positron pairs. In these successive splittings, the quasi--real photons split into electron--positron pairs, which in turn radiate further photons, initiating an electromagnetic cascade. With increasing energy, the number of allowed splittings grows logarithmically, leading to a rapid proliferation of Fock space components.

This cascading dynamics has profound consequences for the quantum state of the system. Each splitting event entangles newly produced degrees of freedom with the rest of the cascade, while Lorentz time dilation renders the relative phases between different Fock components unobservable. As a result, when one traces over unmeasured or inaccessible modes, the reduced density matrix of the observed sector evolves toward a maximally mixed form. The system thus approaches the Maximal Entanglement Limit (MEL), in which the statistical properties of observables are governed primarily by entanglement and combinatorics, rather than by the coherence of the initial radiation field.

The transition from the Weizs\"acker--Williams regime to MEL therefore marks a qualitative change in the nature of high--energy interactions: a crossover from a high occupation number, quasi--classical field to a highly entangled many--body state. In this sense, pair production and cascading provide a concrete dynamical mechanism for the emergence of entropy--dominated behavior at high energies, foreshadowing the analogous role played by parton branching in QCD \cite{Dokshitzer:1977sg,Altarelli:1977zs}.
\smallskip

The Color Glass Condensate (CGC) \cite{McLerran:1993ni,McLerran:1993ka}  (see \cite{Gelis:2010nm} for review) provides an effective description of QCD at small Bjorken \(x\), in which highly occupied gluon modes are treated as classical gauge fields radiated by static color sources \(\rho\). It encodes nonlinear evolution of parton densities at small $x$ and parton saturation \cite{Gribov:1983ivg,Mueller:1985wy} in an effective field theory framework. 
For a fixed configuration of these sources, the gluon field \(A^\mu_{\rm cl}[\rho]\) is a classical solution of the Yang--Mills equations, and particle production from such a field is coherent and Poissonian.
In this strict classical limit, once the sources are specified, the state is effectively pure and carries minimal entropy; in particular, it does not exhibit quantum entanglement among the produced gluons.

Physical observables in the CGC are obtained by averaging over the ensemble of color sources,
\[
\langle \mathcal{O} \rangle_Y
=
\int \mathcal{D}\rho \;
W_Y[\rho]\,
\mathcal{O}\!\left[A^\mu_{\rm cl}(\rho)\right],
\]
where the gauge--invariant weight functional \(W_Y[\rho]\) evolves with rapidity according to the JIMWLK equation \cite{Jalilian-Marian:1997jhx,Iancu:2000hn}, or, at large $N_c$, by the Balitsky-Kovchegov equation \cite{Balitsky:1998ya,Kovchegov:1999yj}.
From the viewpoint of quantum information, this averaging has a clear interpretation: it arises from tracing over fast degrees of freedom that have been integrated out in the construction of the effective theory.
In this sense, the probability functional \(W_Y[\rho]\) carries information about the entanglement between fast (source) and slow (field) modes, see \cite{Kovner:2015hga,Kovner:2018rbf,Duan:2020jkz}.

It is essential, however, to distinguish this effective encoding from a full description of entanglement.
Quantum entanglement is a property of the reduced density matrix and involves phase information and off--diagonal correlations.
By contrast, \(W_Y[\rho]\) is a classical probability distribution.
It does not encode the complete entanglement structure of the underlying QCD wave function, but only those aspects that survive the coarse--graining inherent in the CGC framework.
In particular, entanglement among small--\(x\) gluon modes themselves is largely suppressed in the classical field limit, while correlations between fast and slow modes are retained only in an averaged, probabilistic form.

At the level of a single source configuration, entanglement in CGC is minimal and particle production is Poissonian.
After averaging over sources, fluctuations and entropy reappear, reflecting the typicality of the reduced state obtained after tracing out unobserved degrees of freedom.
The saturation regime corresponds to a situation in which the entropy encoded in \(W_Y[\rho]\) becomes maximal within the available phase space, even though the description in terms of classical fields remains valid.
\smallskip

It is instructive to compare the CGC and MEL frameworks from the viewpoint of statistics. 
 In the CGC framework, particle production is driven by strong classical color fields generated by static color sources. For a fixed configuration of these sources, emissions are independent and coherent, leading naturally to Poisson statistics for produced quanta. 

A geometric (or exponential) multiplicity distribution, in contrast, cannot be obtained as a convolution of Poisson processes in a cascade of emissions. Instead, it arises as a continuous superposition of Poisson distributions with a fluctuating mean,
\begin{equation}
P_n = \int_0^\infty d\lambda \, \frac{e^{-\lambda}\lambda^n}{n!}\, w(\lambda),
\end{equation}
where $w(\lambda)$ is an exponential weight. Such Poisson--Gamma mixtures maximize entropy for a given average multiplicity and therefore provide a natural statistical realization of the MEL. 

This comparison clarifies the conceptual distinction between the two regimes. The CGC corresponds to a situation in which the reduced density matrix remains close to diagonal in a coherent--state basis, leading to Poissonian fluctuations (for fixed sources). The MEL, on the other hand, emerges when quantum entanglement and tracing over inaccessible phases induce large event--by--event fluctuations of effective sources, yielding geometric multiplicity distributions and maximal entropy.

\subsection{Testing the maximal entanglement}

The entropy associated with the dipole multiplicity distribution,
\[
S(Y)
= - \sum_n P_n(Y)\,\ln P_n(Y)
\simeq \ln \langle n(Y)\rangle + \mathcal{O}(1),
\]
can be directly interpreted as the entanglement entropy of the dipole subsystem.
This quantity admits a transparent physical interpretation when related to deep inelastic scattering.
In the dipole picture of DIS, the virtual photon fluctuates into a \(q\bar q\) dipole, which subsequently interacts with the target via its dipole content.
The total DIS cross section, and hence the structure function \(F_2(x,Q^2)\), is proportional to the number of dipoles of size \(r \sim 1/Q\) available at rapidity \(Y\),
\[
F_2(x,Q^2) \;\propto\; \langle n(Y,Q^2)\rangle .
\]

Combining this relation with the expression for the entropy immediately yields
\[
S(x,Q^2)
\;\simeq\;
\ln F_2(x,Q^2) + {\rm const},
\]
establishing a direct connection between the entanglement entropy of the hadronic wave function and experimentally measurable DIS observables.
This relation mirrors Boltzmann's formula \(S=\ln W\), with the role of the number of accessible microstates \(W\) played by the structure function.
\smallskip

At sufficiently large rapidity, nonlinear effects become important and limit the growth of dipole number.
In the dipole formulation, this corresponds to the onset of saturation.
While saturation tames the growth of \(\langle n\rangle\), the system remains close to a maximally entangled state within the available phase space.
The entropy then saturates at a value controlled by the saturation scale \(Q_s(Y)\). 
\smallskip

In summary, the dipole formulation of small-\(x\) QCD provides a concrete realization of how unitary quantum evolution of a pure state leads, through entanglement and coarse graining, to a probabilistic description in terms of parton distributions.
The structure function \(F_2\) thus acquires a clear quantum--informational meaning: it measures the effective number of entangled degrees of freedom resolved by the probe.

\section{Approach to maximal entanglement\\ from first--principles quantum simulations}\label{qsim}

\subsection{MEL from real-time unitary evolution}

An explicit and controlled demonstration of the approach to the Maximal Entanglement Limit (MEL) can be obtained from first--principles quantum simulations of relativistic gauge theories. In a recent series of papers \cite{Florio:2023dke,Florio:2024aix,Florio:2025hoc}, we carried out this program in the Schwinger model \cite{Schwinger:1962tp}, QED in $(1+1)$ dimensions. 

The Schwinger model shares several essential qualitative features with four--dimensional QCD, making it a valuable theoretical laboratory for studying nonperturbative dynamics from first principles. Despite its simplicity, the model exhibits confinement: electric field lines do not spread, and isolated charges are not part of the physical spectrum, closely paralleling color confinement in QCD. 
\smallskip

In fact, in his original 1962 paper \cite{Schwinger:1962tp}, Schwinger wrote: ``{\it Thus, one could anticipate that the known spin-$0$ bosons, for
example, are secondary dynamical manifestations of strongly coupled primary fermion fields and vector gauge fields. This line of thought emphasizes that the
question ``Which particles are fundamental?" is incorrectly formulated. One should ask ``What are the
fundamental fields?"}". This foresight anticipated the emergence of hadrons from fundamental quarks and gluons in QCD.
\smallskip

 Schwinger model also possesses a chiral anomaly, whereby the axial current is not conserved at the quantum level, and this anomaly generates a mass gap in the spectrum, analogous to $\eta'$ mass generation in QCD. Furthermore, the model develops a nonzero chiral condensate providing a clean realization of vacuum structure effects.
 \smallskip

An additional feature that makes the Schwinger model particularly illuminating for the present discussion is its tunable dynamical character. In the massless limit, the model is exactly solvable and integrable: it can be mapped to free massive boson theory. At finite fermion mass, however, integrability is broken, interactions between modes become nontrivial, and the system exhibits quantum chaos. This transition from integrable to chaotic dynamics mirrors, in a controlled setting, the conditions under which entanglement generation and thermalization become efficient. As such, the Schwinger model provides a bridge between analytically tractable field theories and genuinely chaotic quantum systems, allowing one to study how confinement, anomalies, and chiral symmetry breaking interplay with entanglement growth and the approach to the Maximal Entanglement Limit.
\smallskip

In papers \cite{Florio:2023dke,Florio:2024aix,Florio:2025hoc}, the Schwinger model was used to mimic the production of QCD jets. For this purpose, as originally proposed by  
Casher, Kogut and Susskind \cite{Casher:1974vf}, we coupled the theory to two high-energy external sources that propagating back-to-back along the light cone. This electric field produced by the jets is dynamically screened by quark-antiquark pair creation, that resembles fragmentation of jets. In the massless Schwinger model, this real-time process is described by an exact analytical solution found and explored in \cite{Loshaj:2011jx,Kharzeev:2012re,Kharzeev:2013wra}. At finite fermion mass, the model is not integrable, but can be studied numerically with Hamiltonian simulation methods \cite{Florio:2023dke,Florio:2024aix, Florio:2025hoc,Janik:2025bbz}. This is the regime that is most interesting to us from the MEL perspective.
\smallskip

In the Hamiltonian formulation the continuum version of the theory is given by
\be
H =\int dx \left [ \frac{1}{2}E^2+\bar{\psi}(-i\gamma^1\partial_1+g\gamma^1A_1+m)\psi + j_{ext}^1 A_1 \right ], \label{eq:cont_Ham}
\ee
where $\psi$ is the single-flavor two-component fermion field, $A_1$ is the $U(1)$ gauge potential\footnote{We have fixed the temporal gauge $A^0 = 0$.} and $E = F_{01}$ is the electric field strength. $m$ and $g$ are the fermion mass and electric charge, respectively; note that the latter has mass dimension 1 in $(1+1)$ dimensions. The external source $j_{ext}^1$ models two jets propagating back to back along the lightcone:
\begin{equation}
    j_{ext}^1(x,t) = g[\delta(x-t) + \delta(x+t)]\theta(t) \, , \label{eq:jext}
\end{equation}
where we have set the point where the jets first appear as the origin of spacetime.
\smallskip

This theory was simulated using tensor--network and exact diagonalization techniques; I refer the reader to the papers \cite{Florio:2023dke,Florio:2024aix,Florio:2025hoc} for the description and references for the lattice formulation (Kogut-Susskind fermions), map onto a spin chain Hamiltonian (Jordan-Wigner transformation), and the implementation of the real-time Hamiltonian evolution. Our key emphasis here will be the results on entanglement.

\begin{figure}
    \centering
    \includegraphics[width=0.6\textwidth]{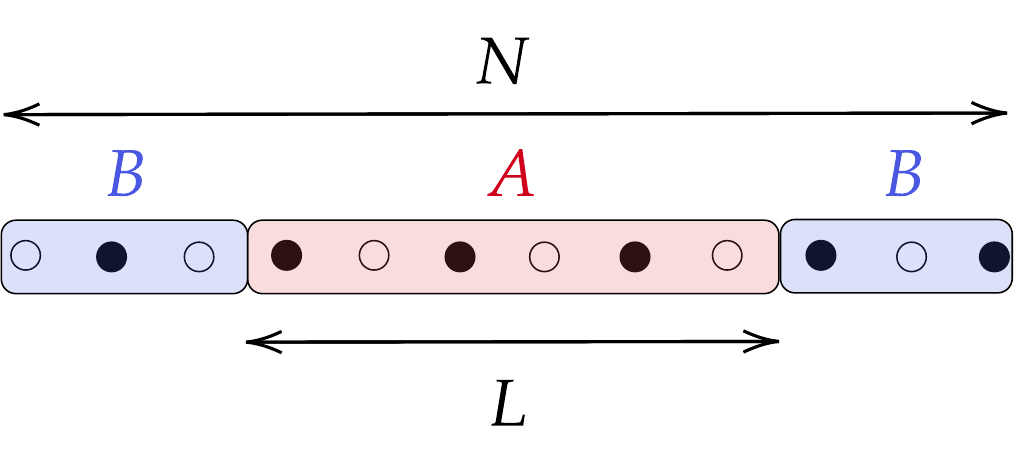}
    \caption{The bipartition of the system used to study the entanglement of the central region of length $L$ (subsystem $A$) with the complement (subsystem $B$). Subsystem $A$ is centered around the center of the system of size $N$. From \cite{Florio:2025hoc}.}
    \label{fig:AB_bipartition}
\end{figure}

In particular, we are interested in the real-time dynamics of the entanglement entropy of a subsystem consisting of  an interval at the center of the system, with its complement; this bipartition is displayed in Fig.~\ref{fig:AB_bipartition}. The entanglement entropy is defined as
\begin{align}
    S(t) = -\tr_A(\rho_A(t)\log \rho_A(t))  \, , 
\end{align}
where $\rho_A(t) = \tr_B\rho(t)$, $\rho(t) \equiv |\psi(t)\rangle\langle\psi(t)|$ and $\tr_i$ denotes tracing over the degrees of freedom of subsystem $i = A,B$. We find the spectrum of the reduced density matrix $\rho_A$, also known as Schmidt eigenvalues, $\{\lambda_i\}$, from a Matrix Product State (MPS) representation of the quantum state, see \cite{Florio:2025hoc} for details.

\begin{figure}
    \centering
    \includegraphics[width=0.6\linewidth]{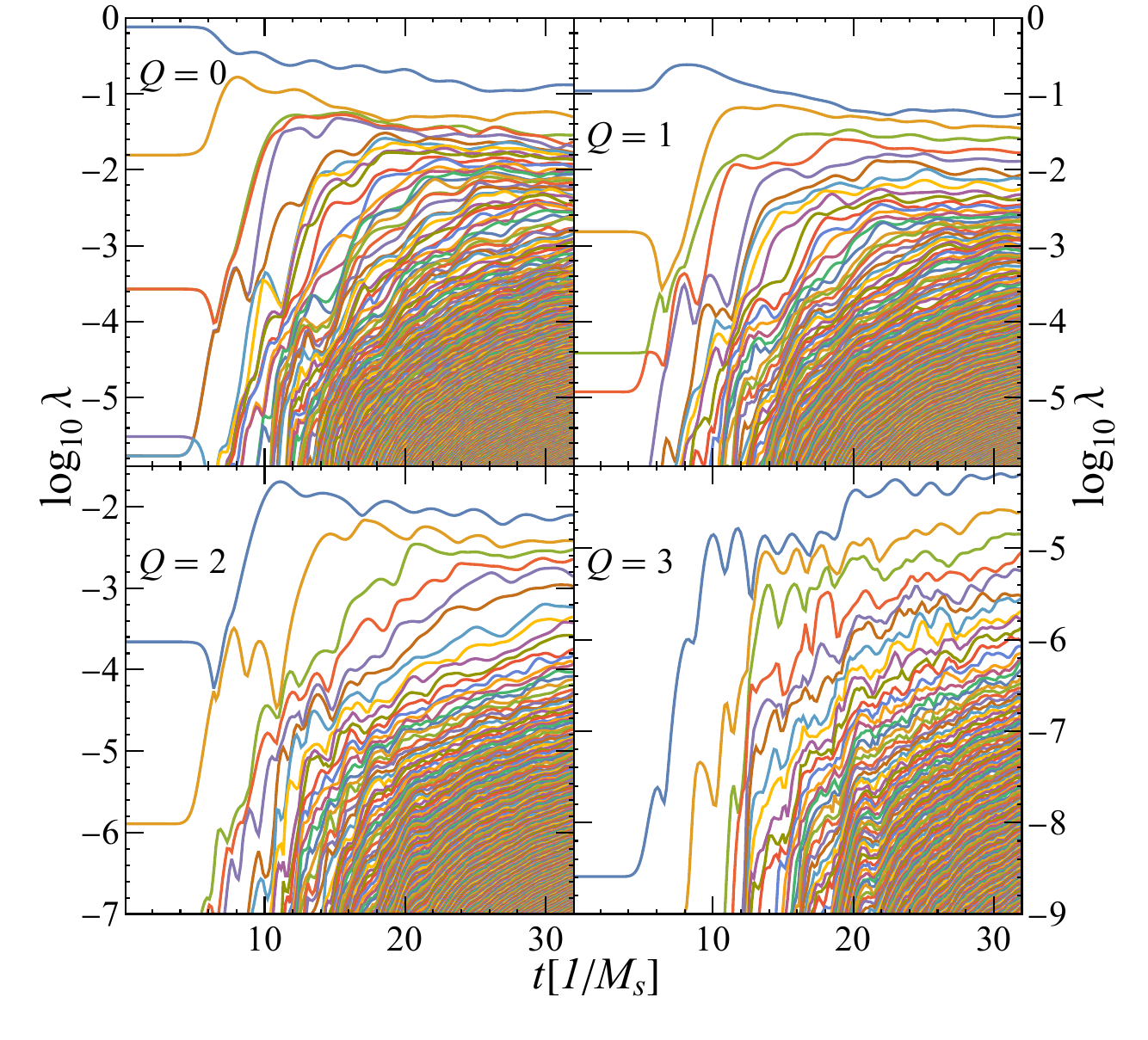}
    \caption{Entanglement spectrum in four different charge $Q$ sectors for a subsystem of size $L=12$ as a function of time (logarithms of eigenvalues are shown for legibility). Total size of the system is $N=100$, fermion coupling and mass are $g = 0.5/a, m = 0.5 g$. From \cite{Florio:2025hoc}.}
    \label{fig:ent_spectrum}
\end{figure}

The entanglement spectrum in a subsystem of size $L=12$ is shown in Fig.~\ref{fig:ent_spectrum};  we consider the system with parameters $g=0.5/a, m=0.5g$ and $N=100$. Because of the presence of a conserved electric charge $Q$, each eigenstate of the reduced density matrix has a particular charge. In Fig.~\ref{fig:ent_spectrum} we organize the eigenvalues into 4 different panels by charge sector. 

In each charge sector we find that at early times, very few Schmidt states contribute to the reduced density matrix, reflecting that the subsystem is close to a pure state. However, as time evolves, many more Schmidt states emerge and the subsystem moves towards the MEL! 

This is in accord with our general discussion in Section \ref{geomh}:  Schwinger model is a relativistic quantum field theory with a very high-dimensional Hilbert space, and once enough energy is supplied from external sources to sample this space, the density matrix of a small subsystem approaches the MEL.

\begin{figure}
    \centering
    \includegraphics[width=0.5\linewidth]{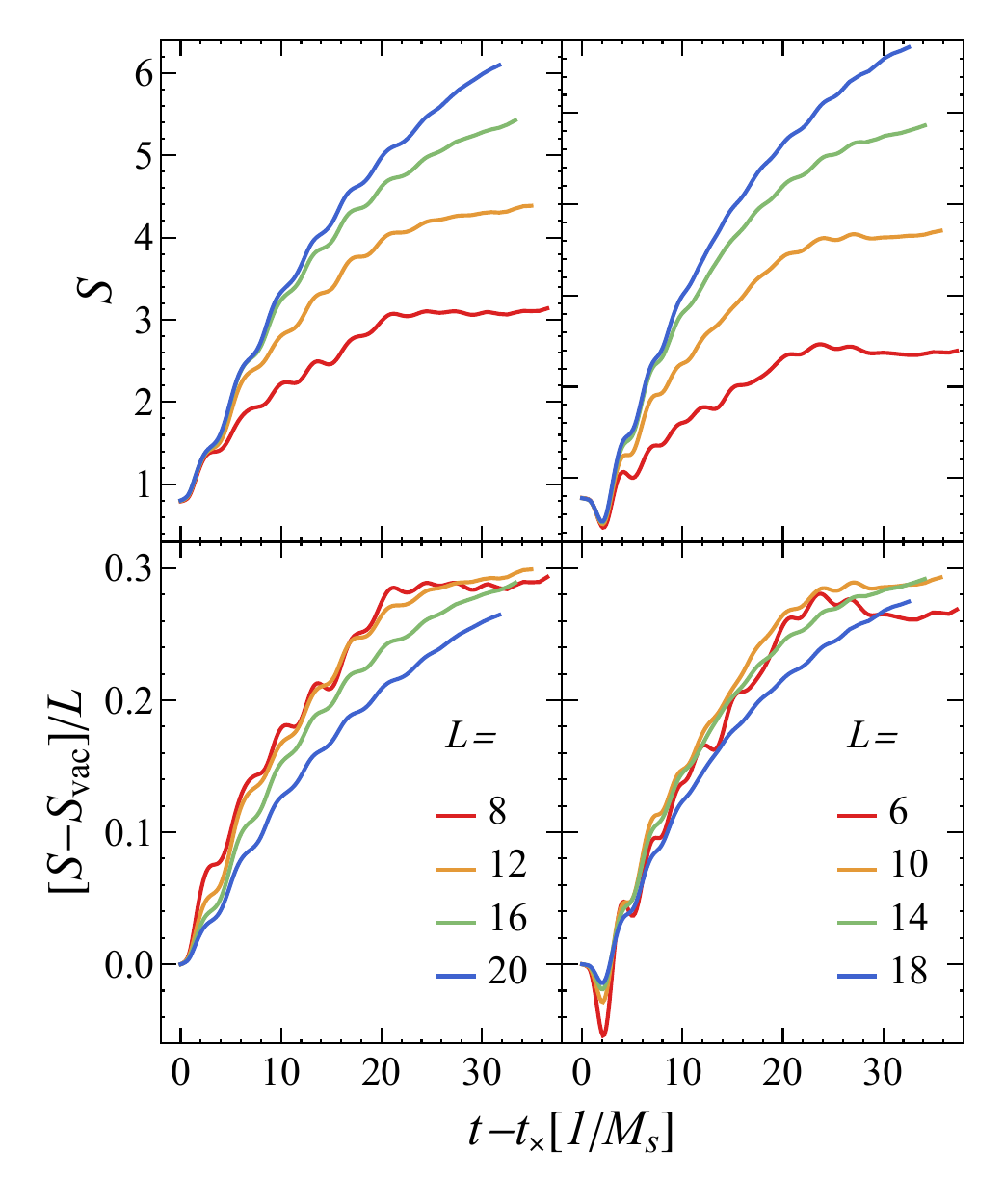}
    \caption{Top panel: time dependence of entanglement entropy $S(L)$ (times are shifted so that the jets cross the subsystem boundary at $t-t_\times=0$). Matching of the initial shape of the curves is the signature of an area law scaling of the entanglement entropy, expected for the ground state of a gapped theory.  Bottom panel: vacuum--subtracted entanglement entropy per unit length, with the time shifted same as above. Different subsystem size curves reach a plateau at late times at the same value of entropy density, signifying the volume law of entanglement that signals thermalization. System parameters are $m=0.5 g, g=0.5/a$; values of $L$ are separated on the right and on the left panels due to staggering effects. From \cite{Florio:2025hoc}.}
    \label{fig:S_area_vol_law}
\end{figure}

From the entanglement spectrum, it is straightforward to compute entanglement entropy as 
\begin{equation}
    S(t) = -\sum_i \lambda_i \log\lambda_i. 
\end{equation}
Let us discuss the real-time dependence of this quantity. The external charges cross the boundary between subsystems $A$ and $B$ at $t_\times = L/2$. This is the moment when the entanglement entropy starts changing. Offsetting by the jet arrival times, we find that the initial time dependence of entanglement entropy follows a universal behavior, see the top panel of Fig.~\ref{fig:S_area_vol_law}. It is a signature of the area law of entanglement, characteristic for ground state and weakly entangled states of gapped theories~\cite{Eisert:2008ur} (in lattice formulation, this behavior is different for even and odd $L/2$ due to the staggering effects). 

After an intermediate stage of growth, the entanglement entropy saturates at late times. By subtracting the vacuum value of the entropy and rescaling by the subsystem size, we find that the entanglement entropy follows a volume law at late times, as shown in the bottom panel of Fig.~\ref{fig:S_area_vol_law}. As volume law is expected for a thermal state~\cite{Bianchi:2021aui, Nakagawa:2017yiw}, this is further evidence of thermalization.
\smallskip

It is now natural to ask how close, quantitatively, the quantum state of the produced system is to a thermal state. In order to make a full comparison, we focus on a subsystem $A$ of size $L$ of the jet system (see Fig. \ref{fig:AB_bipartition}) described in terms of a reduced density matrix $\rho_A$. We then measure the similarity of this reduced density matrix to thermal states, characterized by density matrices $\rho_\beta$, and determine a temperature by selecting the thermal state closest to $\rho_A$. To measure the proximity of $\rho_A$ and $\rho_\beta$,  we compute the \textit{overlap} between the two mixed states:
\begin{equation}
    f(\rho_A,\rho_\beta) \equiv \frac{\Tr (\rho_A \,\rho_\beta)}{\sqrt{\Tr\,\rho_A^2  \, \Tr\, \rho_\beta^2}}\,. \label{eq:overlap_definition}
\end{equation}
The overlap is equal to unity if $\rho_A=\rho_\beta$ and can be computed with tensor network methods, see \cite{Florio:2025hoc} for details.  
\begin{figure}
    \centering
    \includegraphics[width=0.6\linewidth]{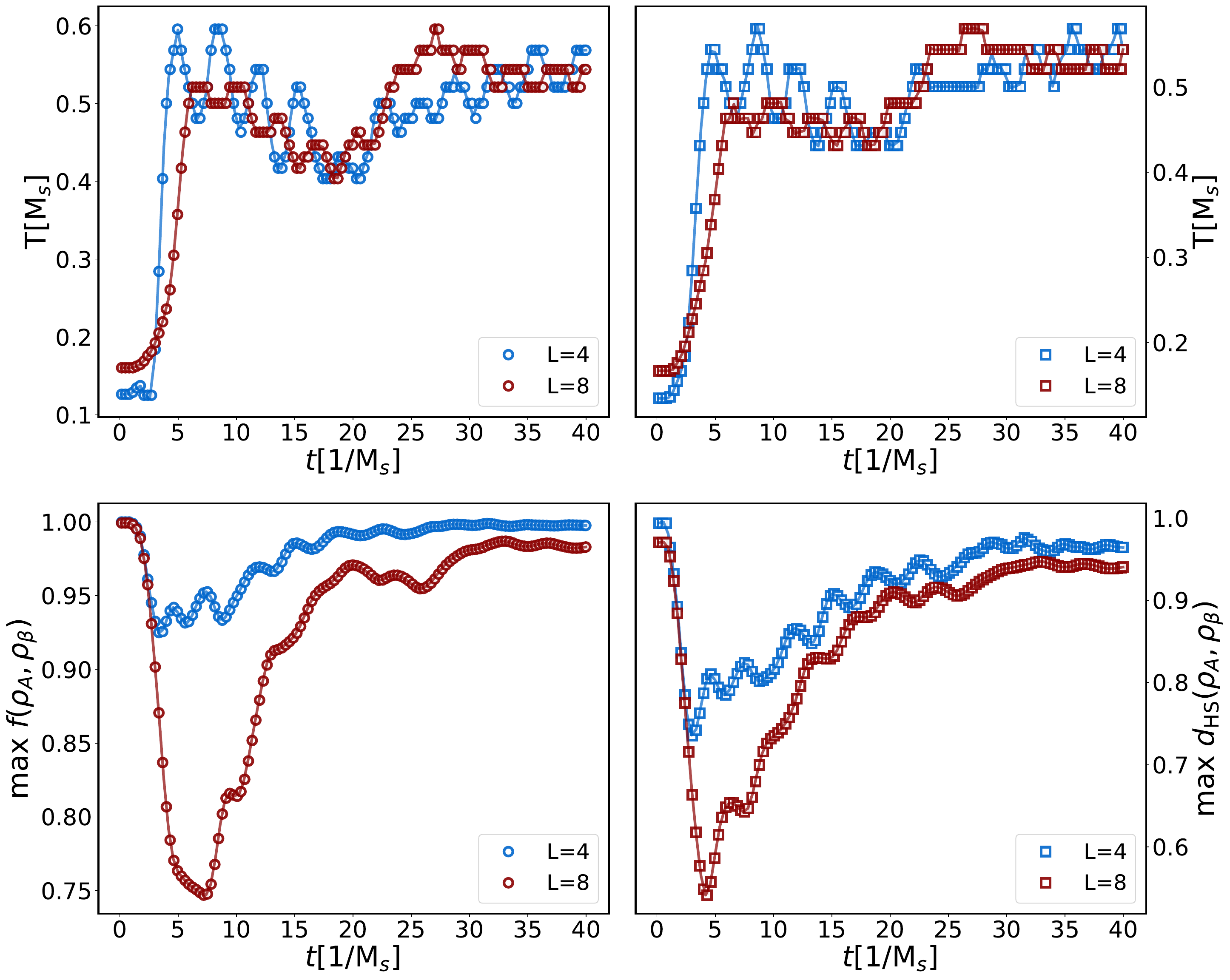}
    \caption{Top panel: The temperature as a function of time extracted from the maximal normalized overlap (right) and minimal HS distance (left) for subintervals $L=4,8$ centered in the middle. Bottom panel: the value of the metric at the extremum shown in the top panel. For the thermal states we used $N_{\rm thermal}=12$ and traced down to $L$. From \cite{Florio:2025hoc}.}
    \label{fig:overlaptime}
\end{figure}
As a second measure, instead of using the trace distance we consider the (square root of the) Hilbert-Schmidt (HS) distance~\cite{Vedral:1997hd}
\begin{equation}
    D_\text{HS}(\rho_A,\rho_\beta)=\sqrt{\Tr[(\rho_A-\rho_\beta)^2]}.
\end{equation}
The HS distance bounds the trace distance from above \cite{Coles:2019wyr}. It is convenient, instead of using the HS distance, to use its complement
\begin{equation}
    d_\text{HS}(\rho_A,\rho_\beta)=1-D_\text{HS}(\rho_A,\rho_\beta),
\end{equation}
that approaches 1 when $\rho_A\simeq \rho_\beta$.
\begin{figure}
    \centering
    \includegraphics[width=0.6\linewidth]{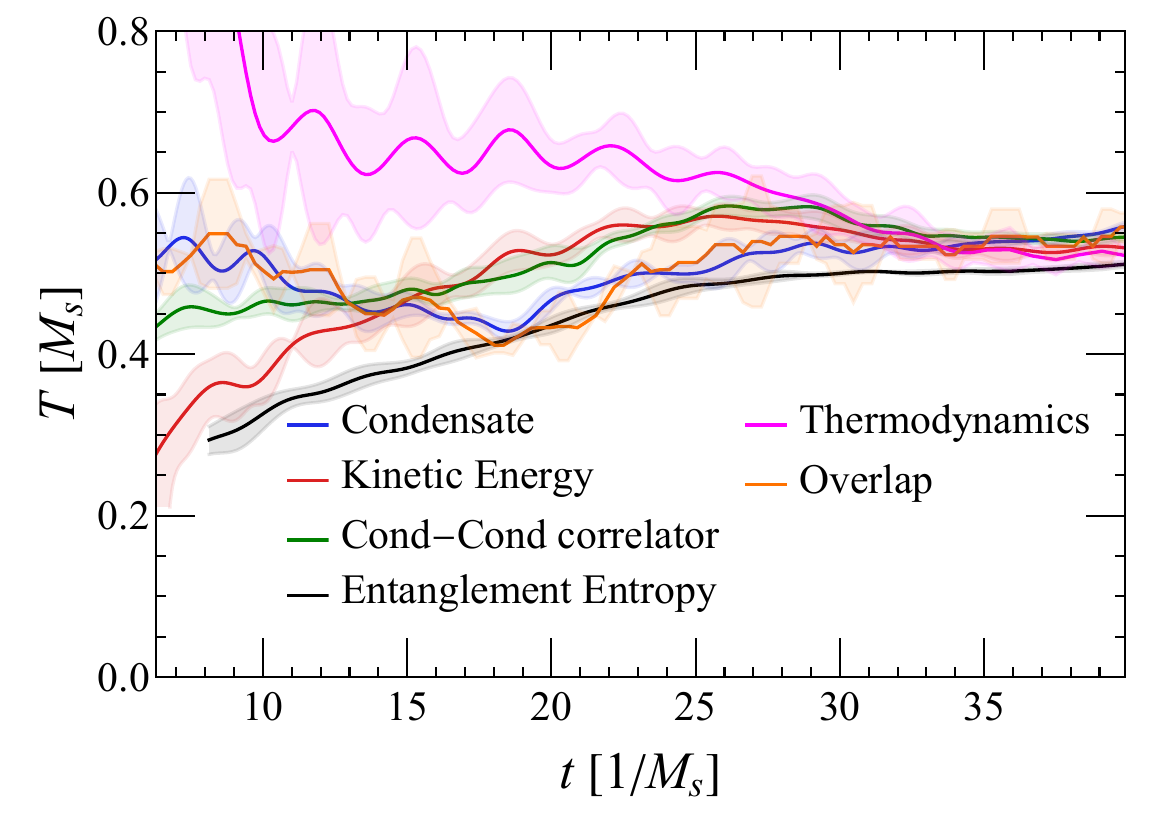}
    \caption{Effective temperature $T$ found from the local observables, entanglement entropy, thermodynamic relation $s=(\epsilon+P)/T$ ($\epsilon$ and $P$ are the energy density and pressure) and maximizing the overlap between the reduced density matrix and a thermal density matrix.  Fermion coupling and mass are $g=0.5/a$,  $m=0.5\,g$. Temperature $T$ and time $t$ are defined in terms of the meson mass $M_S$. From \cite{Florio:2025hoc}.}
    \label{fig:thermalization_m025}
\end{figure}
Fig.~\ref{fig:overlaptime}, bottom panel, shows the maximum value of overlap and complement HS distance as a function of time (for a system with $g=0.5/a$, $m=0.5g$ and $N_{\rm jet}=100$). We see that at early times, before the sources are turned on, the overlap is largest for thermal state with largest $\beta$ ($\beta=10$) since the system is still in the vacuum state. Upon switching on the sources, the overlap with thermal states drops significantly since we are far from equilibrium. As time evolves, the two metrics start growing, and at late times approach unity. This shows that $\rho_A$ approaches a thermal state.
\smallskip

We can also extract the temperature can be extracted from the $\rho_\beta$ state that maximizes the metric, see Fig.~\ref{fig:overlaptime}, top panel. Alternatively, we can extract this temperature by comparing the computed values of various observables to their thermal expectation values. While for a given observable one can always find an equivalent temperature defined in this way, in general these temperature will be different for different observables. 

However, we find that at late times all local observables, and correlation functions, yield the same universal temperature, as expected for a thermal state -- see Fig.~\ref{fig:thermalization_m025}. Moreover, this temperature coincides with the one extracted from the density matrix overlap with the thermal one (the temperature extracted from maximal overlaps for $L=4,8$ is labeled as ``overlap'' in Fig.~ \ref{fig:thermalization_m025}). 
\smallskip

The simulations described above establish, from first principles, that maximal entanglement and thermal reduced density matrices need not rely on ergodicity or stochastic assumptions. Instead, they emerge naturally from quantum entanglement generated by real-time unitary evolution and subsequent tracing over inaccessible degrees of freedom. Therefore, these results provide a microscopic underpinning for the MEL advocated in these lectures, and offer a treatable prototype for understanding analogous phenomena in high--energy QCD, where direct simulations currently remain out of reach.

\subsection{MEL in the quantum breakup of confining strings}

A particularly interesting realization of the Maximal Entanglement Limit (MEL) in a {\it static} setting is provided by recent quantum simulation studies of static confining strings in Schwinger model \cite{Grieninger:2025rdi}. In this work, the system is initialized in a highly excited but pure quantum state corresponding to a static electric string stretched between external charges. Although at small spatial separation between the charges the state is coherent and low--entropy, 
the increase in the inter-charge distance leads to string breaking through the production of particle--antiparticle pairs, accompanied by the entanglement generation.
\smallskip

We considered two types of partitions: a subsystem of length $L$ centered in the middle of the lattice (in between the static charges), and a half-chain bipartition separating the system into left and right halves. The first one probes the spatial extent of entanglement inside the flux tube, while the second captures the global buildup of correlations as the external charges are separated. 

\begin{figure}[ht!]
    \centering
    \includegraphics[width=0.45\linewidth]{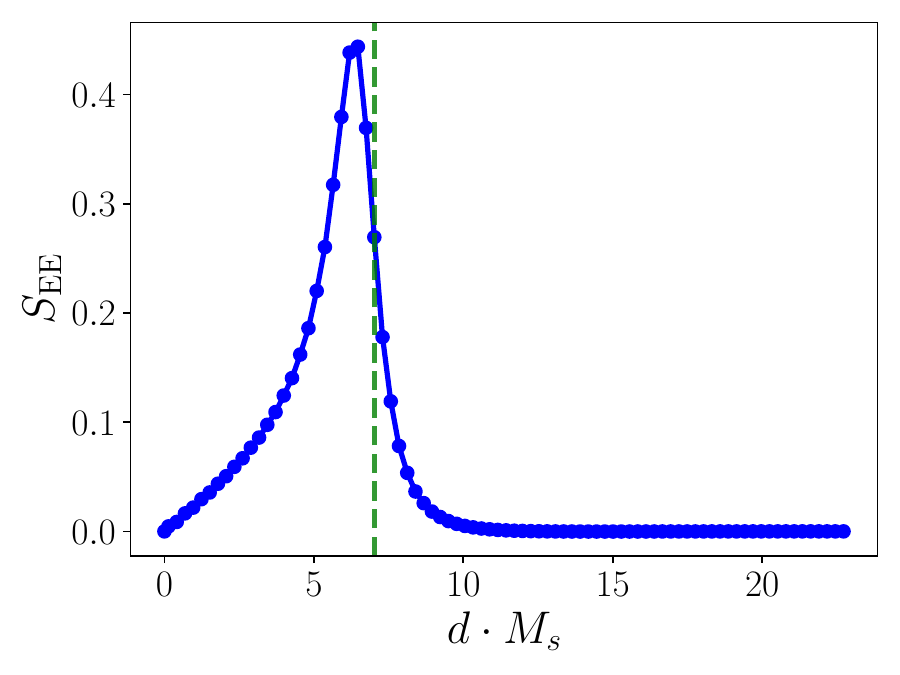}
    \includegraphics[width=0.45\linewidth]{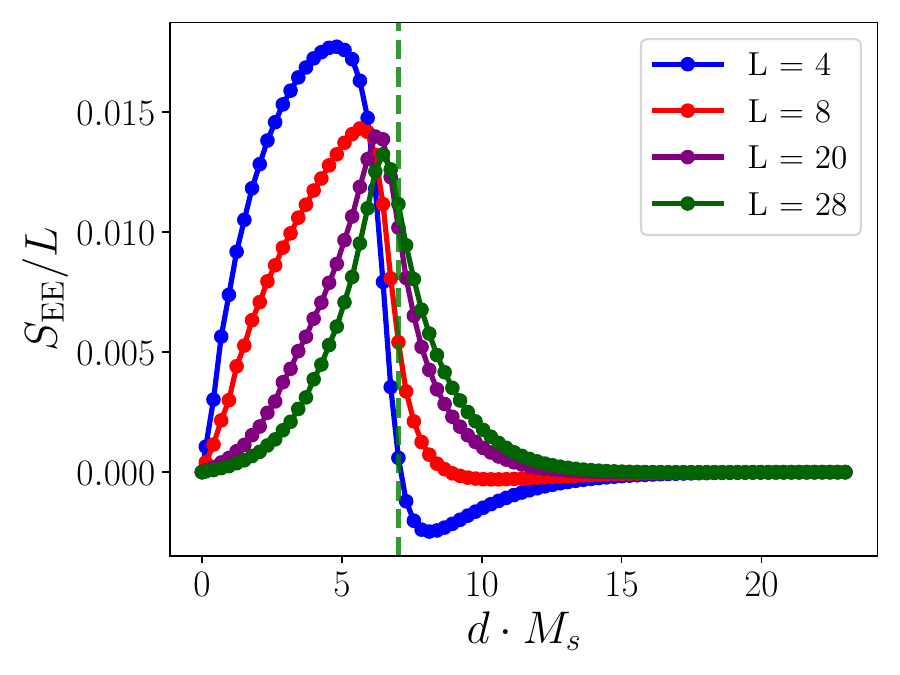}    
    \caption{Entanglement entropy as a function of the charge separation $d \cdot M_s$ for different partitions of the system. Left: Half-chain entropy capturing the total entanglement between the left and right halves of the lattice. Right: Entanglement entropy of centered subsystems of length $L$, measured relative to the vacuum -- hence the region of negative $S_{EE}$. From \cite{Grieninger:2025rdi}.}
    \label{entropy}
\end{figure}

The half-chain entropy, shown in the top panel of Figure \ref{entropy}, displays a striking non-monotonic behavior: it rises sharply with separation, reaches a peak near the critical distance at which the string breaks up, and then returns toward its vacuum value at large $d$.   This peak provides a clear signature of the confining string and its decay, consistent with the tensor-network results of \cite{Buyens:2015tea}.

Fig. \ref{entropy} (right) shows the EE for various subsystem sizes $L$, measured relative to the vacuum. For small separations, the entanglement is weak; however,  as the distance $d$ increases, the EE rapidly grows, reflecting the buildup of quantum correlations within the confining flux tube.  

\begin{figure}[ht!]
     \centering
     \includegraphics[width=0.45\linewidth]{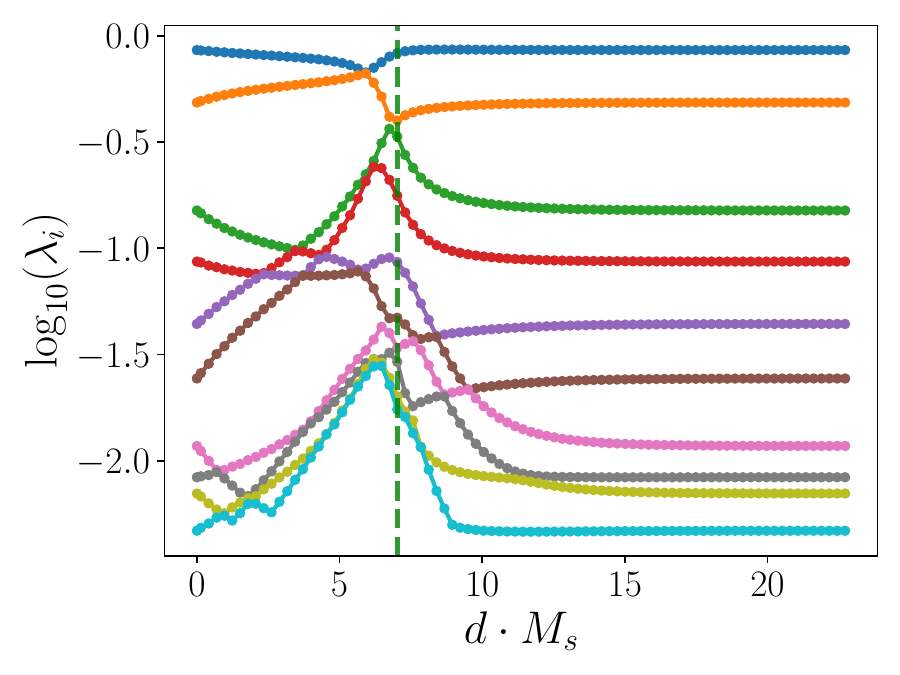}   
          \includegraphics[width=0.45\linewidth]{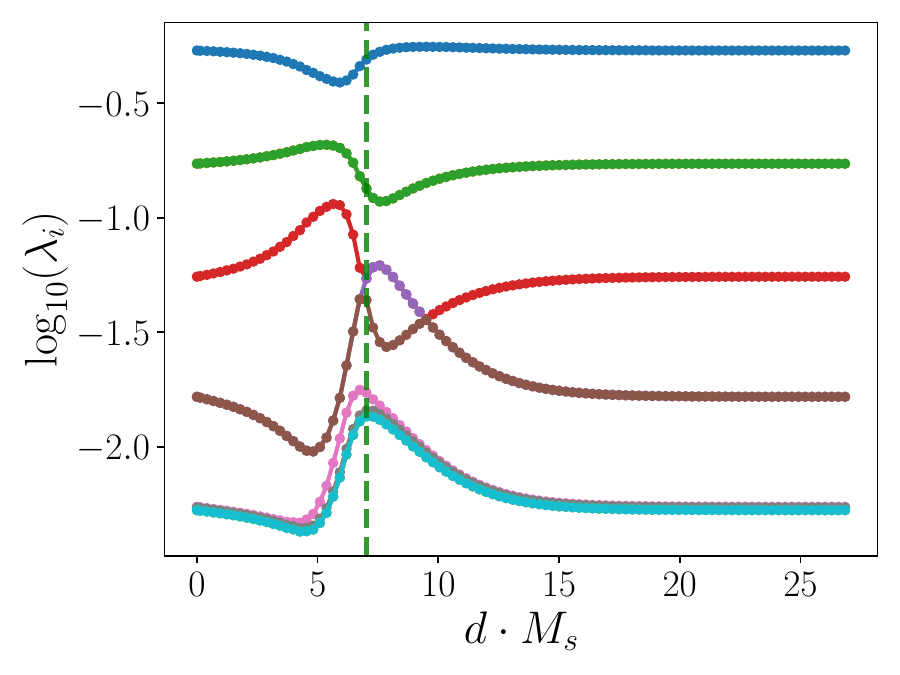}   
     \caption{Left: entanglement spectrum for half of the system which yields the bipartite entanglement entropy shown in Fig. \ref{entropy}.
     Right: Entanglement spectrum of the reduced density matrix $\rho_A$ for a centered $L=32$ subsystem, shown as the logarithm of the ten largest eigenvalues versus the separation between the external charges $d \cdot M_s$. From \cite{Grieninger:2025rdi}.}
     \label{spectrum}
     \end{figure}

Beyond  $d_c$, the entanglement rapidly decreases as the flux tube fragments into two separated meson-like states\footnote{Note that we show the difference between the EE and its value in the vacuum -- so the negative value of $S_{EE}$ observed for the smallest size $L=4$ of the subsystem indicates that the entanglement entropy becomes smaller than its vacuum value.}.
This rapid growth of the EE results from a dramatic rearrangement of the entanglement spectrum of the system, as shown in Fig. \ref{spectrum}. As the string is about to break, the Schmidt eigenvalues get closer, and start to resemble the eigenvalues of the thermal density matrix -- the behavior that we have encountered in the previous section.
\smallskip

\begin{figure*}[ht!]
     \centering
     \includegraphics[width=0.32\linewidth]{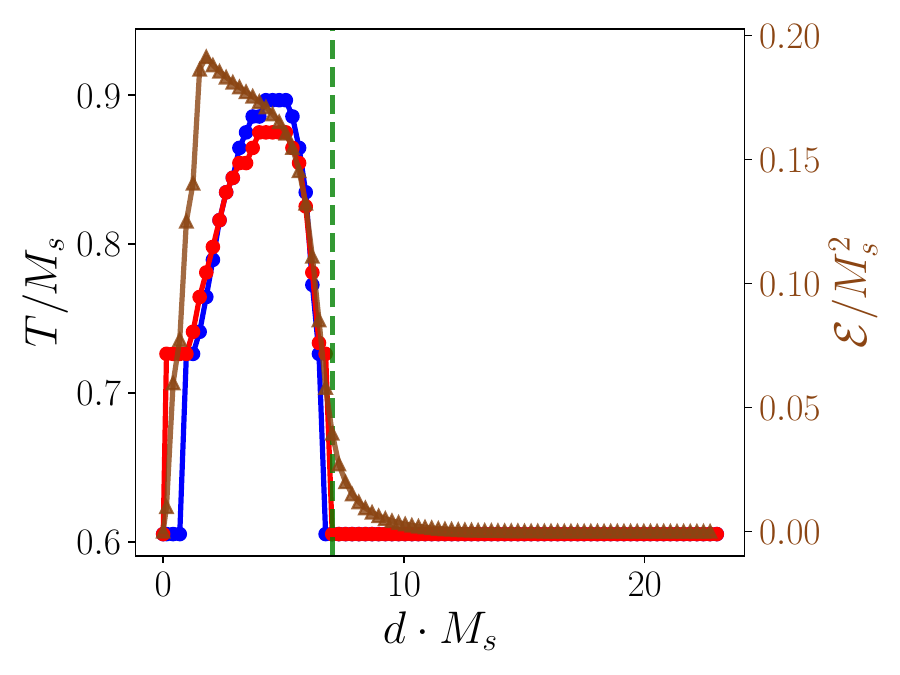}     \includegraphics[width=0.32\linewidth]{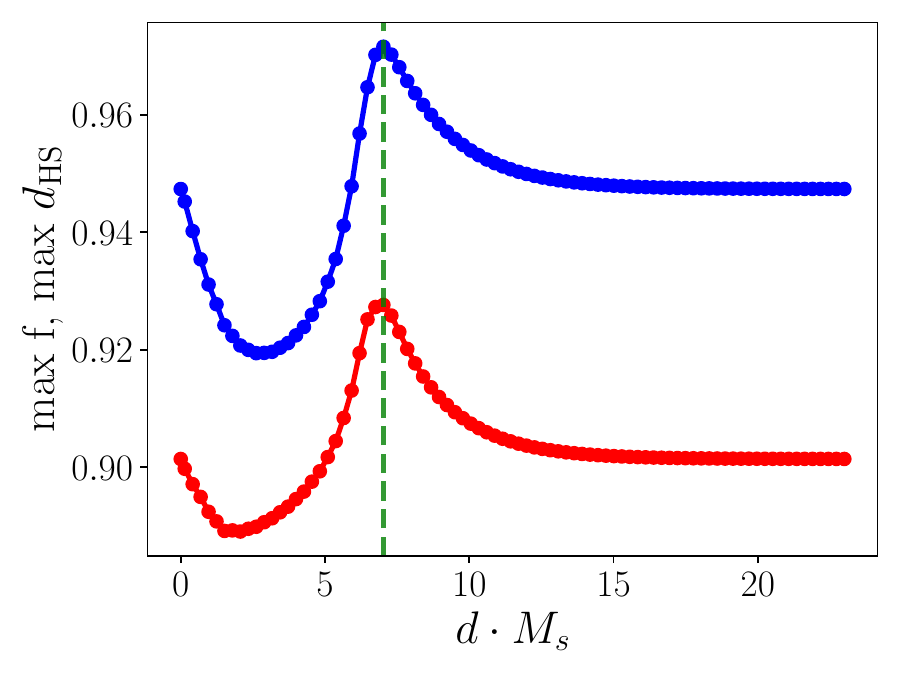}
     \includegraphics[width=0.32\linewidth]{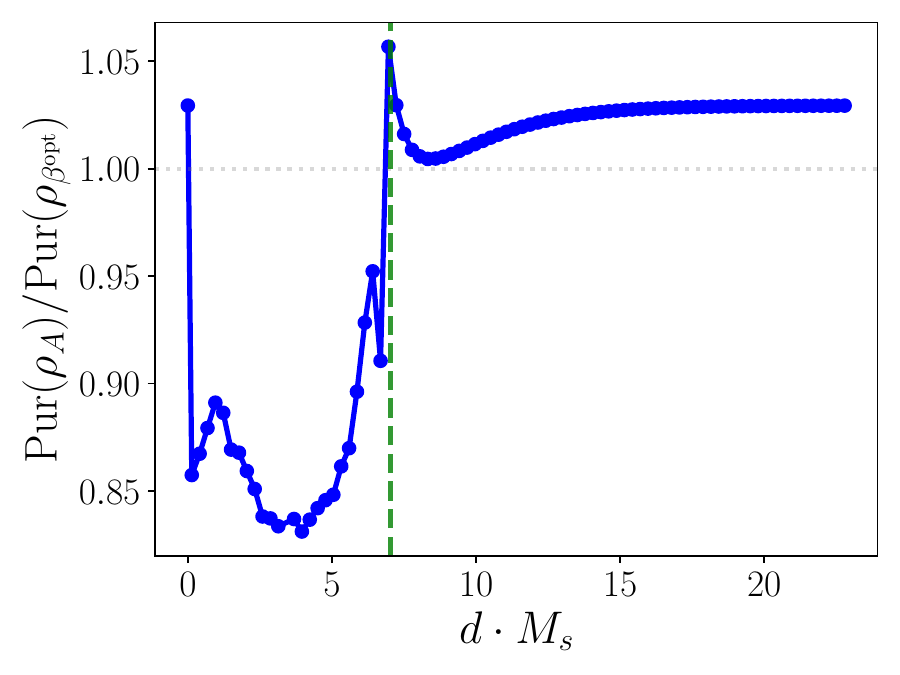}    
     \caption{Subsystem of size $L=12$. Blue dots: Temperature extracted from the maximal normalized overlap (left) and the corresponding maximal normalized overlap as a function of $d\cdot M_s$ (middle). Red squares: Temperature from minimal Hilbert-Schmidt distance (left) and the corresponding complement of Hilbert-Schmidt distance at its maximum value as a function of $d\cdot M_s$ (middle). For comparison, the energy density is added (brown triangles). The right plot shows the ratio of the purity of the reduced density matrix to the purity of the thermal density matrix at maximal normalized overlap. From \cite{Grieninger:2025rdi}.}
     \label{overlap12}
     \end{figure*}

As before, we can quantify this similarity by computing the overlaps between the density matrix of the reduced subsystem and the thermal one by using the overlap measures described in the previous section. As can be seen from Fig. \ref{overlap12} (middle panel), both overlap and Hilbert-Schmidt measures indicate that the reduced subsystem is nearly indistinguishable from a thermal ensemble at the critical distance at which the string breaks up.  This result confirms that the flattened entanglement spectrum presented in Fig. \ref{spectrum} indeed corresponds to an effectively thermal mixed state. 
\smallskip

As can also be seen from Fig. \ref{overlap12} (left panel), the effective temperature increases with the separation between the external charges and reaches a pronounced maximum near the critical distance $d_c \cdot M_s \simeq  7$, where the string breaks. Note that as soon as the string breaks, the temperature reaches its vacuum value, even though the transition for the energy density is more smooth (presumably due to the residual interaction between the produced mesons). It is important to keep in mind that an effective temperature of the quantum vacuum is different from zero \cite{Florio:2025hoc}. 

The overlap measures shown in the middle panel peak in the same region, demonstrating that the similarity between $\rho_A$ and $\rho_\beta$ is strongest at the critical distance $d_c$.  In Fig. \ref{overlap12} (right panel), we also show the ratio of the purities Pur$(\rho)\equiv {\rm Tr} (\rho^2)$ for the subsystem of size $L$ and the thermal system of the same size. One can see that slightly above the string-breaking distance this ratio approaches unity.
\smallskip

These results imply that the maximal entanglement and effective thermal behavior are not limited to weakly coupled  branching processes. They can also arise in strongly coupled, confining dynamics that admit a string description. The quantum breakup of confining strings thus provides a complementary demonstration of the MEL.

\subsection{Maximal entanglement erases long-distance confinement}

The emergence of the MEL in the fragmentation of a static confining string admits a simple and natural interpretation in terms of the growth of the accessible Hilbert space within a fixed energy shell, as discussed in section \ref{canons}. In a confining theory, the energy stored in the flux tube increases linearly with the separation $L$ of the external charges,
$$
E(L) \simeq \sigma\, L ,
$$
where $\sigma$ is the string tension. As the string stretches, the total energy available for particle production and excitations grows, leading to a rapid increase in the dimension of the Hilbert space compatible with this energy.

Within a canonical energy shell, the dimension of accessible Hilbert space $\dim \mathcal{H}(E)$ rapidly grows as $\dim \mathcal{H}(E) = e^{S(E)}$, where $S(E)$ is the microcanonical entropy associated with the shell at total energy $E$. Indeed, by definition, this entropy is given by Boltzmann's formula 
$$
S(E) \;\equiv\; \ln \Omega(E),
$$
where $\Omega(E)$ is the number of quantum states with energies in a narrow window $[E,E+\delta E]$. Equivalently, $\Omega(E)$ is the dimension of the Hilbert subspace $\mathcal{H}(E)$ spanned by these states,
$$
\dim \mathcal{H}(E) = \Omega(E) = e^{S(E)} .
$$

This exponential growth has profound implications for entanglement. For the bipartition of the system into a subsystem $A$ and its complement $\bar A$ considered above, the full state should be chosen as a typical pure state within the energy shell $\mathcal{H}(E)$. Page's theorem states that, provided $\mathcal{H}_{\bar A} \gg \dim \mathcal{H}_A$, the reduced density matrix
\begin{equation}
\rho_A = \mathrm{Tr}_{\bar A}\, |\Psi\rangle\langle \Psi|
\end{equation}
is, with overwhelming probability, exponentially close to the maximally mixed (thermal) state on $\mathcal{H}_A$, up to corrections suppressed by the ratio $\dim \mathcal{H}_A / \dim \mathcal{H}_{\bar A}$. 
\smallskip

As a result, as the external charges separate, the density matrix describing the string approaches the maximally entangled form. When this happens, the long-distance information describing confinement gets lost, and the string breaks. The MEL thus emerges as a universal mechanism for the screening of confinement and hadronization. 

In a nutshell, {\it maximal entanglement erases long-distance confinement}. This means that confining strings cannot stretch beyond the critical distance at which the entanglement becomes maximal. 
The MEL may thus be viewed as the quantum-information origin of ``soft confinement" underlying the local parton-hadron duality \cite{Azimov:1984np}, see related discussion in \cite{Dokshitzer:2004ie}.

\subsection{Maximal entanglement produces thermally equilibrated hadrons}

The presented results imply that hadrons resulting from the fragmentation of confining strings are {\it born} in thermal equilibrium, as opposed to the equilibrium reached through final-state interactions. This mechanism was advocated previously in a number of model approaches where thermality appears due to the emergence of an effective event horizon in high energy collisions \cite{Kharzeev:2005iz,Kharzeev:2006zm,Castorina:2007eb,Lin:2009pn,Chesler:2010bi,Castorina:2014fna,Berges:2017zws,Grieninger:2023ehb,Khakimov:2023emy}. 

Basing on the results presented here, it appears that the apparent thermal behavior of the produced hadrons is rooted in the geometry of Hilbert space. Because maximal entanglement appears naturally on the event horizons, geometry of high-dimensional Hilbert space indeed emerges as the underlying reason of the puzzling phenomenological success of statistical hadronization models in high energy collisions \cite{Becattini:1995if,Becattini:2010sk,Andronic:2017pug}, including $e^+e^-$ annihilation into jets, where most of the hadrons produced in jet fragmentation are separated by space-like intervals and thus cannot interact.

\section{Experimental Tests of MEL in High-Energy Interactions}

The concept developed in the previous section leads to a striking and experimentally testable prediction: high-energy hadronic states probed in inclusive processes behave as maximally entangled quantum systems. In close analogy with Boltzmann's relation between entropy and the number of accessible microstates,
\[
S = k_B \ln \Omega ,
\]
it was proposed in Ref.~\cite{Kharzeev:2017qzs} that the entanglement entropy of a hadronic state probed in deep inelastic scattering is directly related to the logarithm of the measured structure function. In this Maximal Entanglement Limit (MEL) correspondence, the role of $\Omega$ is played by the effective number of partonic configurations resolved at a given resolution scale, while the entropy quantifies the entanglement between the probed degrees of freedom and the unobserved ones (i.e. the phases of the Fock components, as dicussed in the previous section).

\subsection{MEL in Deep Inelastic Scattering}

As discussed in section \ref{opm}, in deep inelastic scattering, a virtual photon probes the hadron on a light-cone time scale $\Delta x^+ \sim 1/q^-$. The reduced density matrix relevant for the measurement is obtained by tracing over the unobserved phases \cite{Kharzeev:2021nzh}, leading to an entanglement entropy
\[
S_{\mathrm{DIS}} = -\Tr\left(\rho_{\mathrm{red}} \ln \rho_{\mathrm{red}}\right).
\]
It was argued \cite{Kharzeev:2017qzs,Kharzeev:2021nzh} that, at sufficiently small $x$ where the gluons dominate DIS, this entropy is well approximated by 
\[
S_{\mathrm{DIS}} \simeq \ln \left[ x\,G(x,Q^2) \right],
\]
where $x\,G(x,Q^2)$ is the gluon distribution function. 
\begin{figure}
\begin{center}
     \includegraphics[width=.6\textwidth]{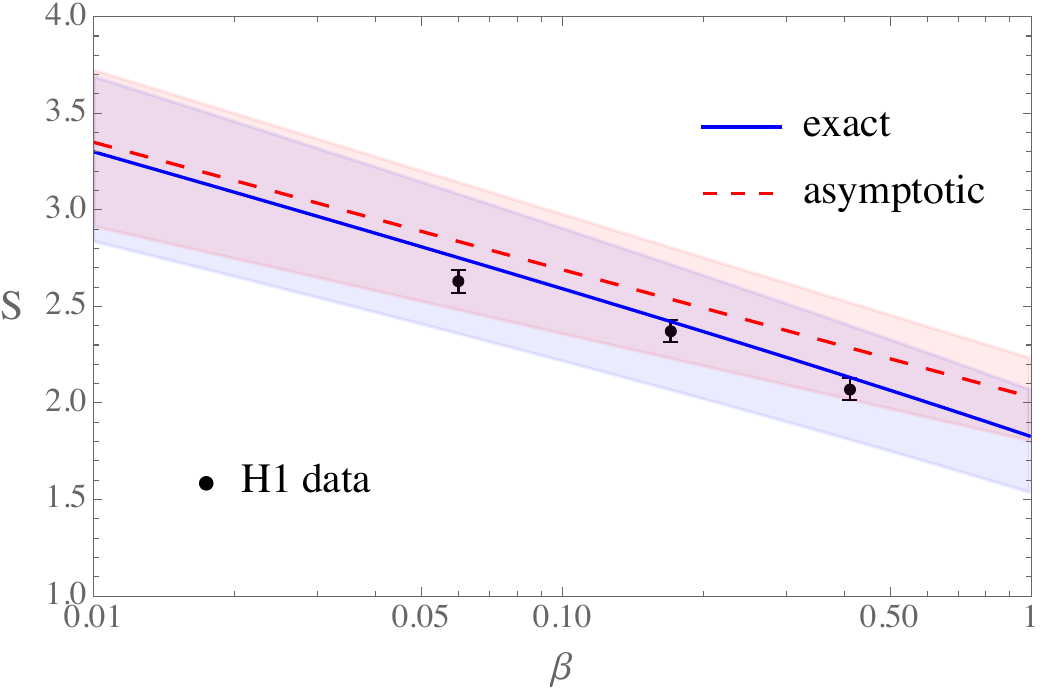}
     \end{center}
     \caption{Comparison between the logarithm of diffractive parton distribution functions (PDF) and hadron entropy as a function of $\beta=x/x_\mathbb{P}$, where $x_\mathbb{P}$ is the fraction of the proton's momentum carried by the Pomeron. H1 data~\cite{H1:1998xpp} for the hadron entropy computed from the multiplicity distributions are shown (statistical and systematic uncertainty are added in quadrature and presented as error bars). The theoretical uncertainty bands correspond to PDF and its scale uncertainty added in quadrature, where the scale uncertainty is obtained from the variation of the factorization scale of the leading order diffractive PDFs in the range $Q \to [Q/2, 2Q]$. From \cite{Hentschinski:2023izh}.
     }
     \label{fig:entropy_vsdata1}
 \end{figure}

This relation mirrors Boltzmann's formula and implies that the hadronic wavefunction probed at small $x$ approaches a state of maximal entanglement. It is based on the explicit solution of a QCD evolution equation in the high-energy limit \cite{Kharzeev:2017qzs}. However, just as Boltzmann limit does not depend on details of microscopic dynamics once the equilibration is reached, this formula is expected to apply irrespectively of the details of the QCD evolution at sufficiently high energy. 
\smallskip
\begin{figure*}
    \centering     \includegraphics[width=.7\textwidth]{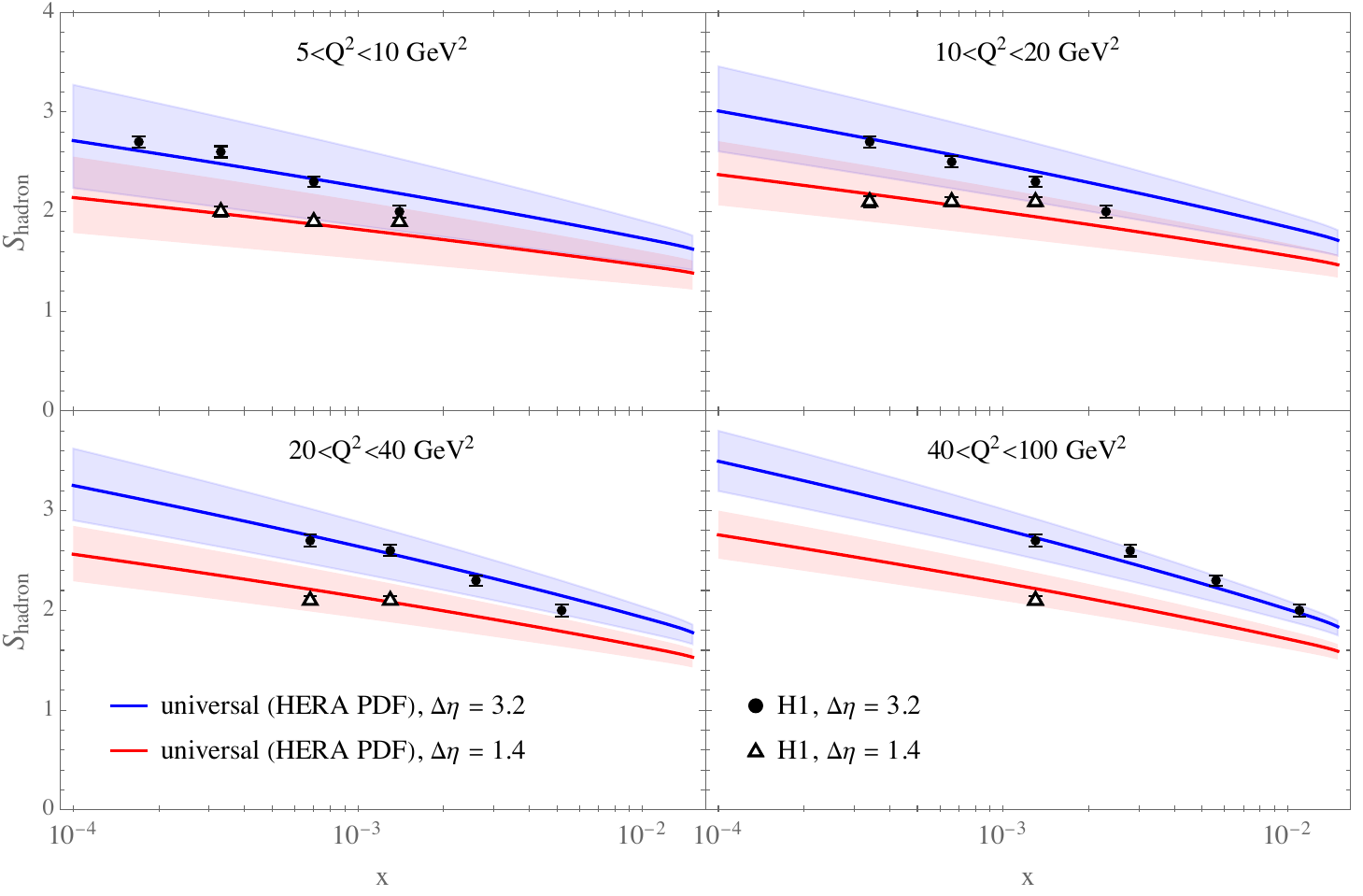}
    \caption{Hadronic entropy vs the logarithm of the leading order HERA PDFs at different momentum transfer $Q^2$ in different rapidity windows. From \cite{Hentschinski:2024gaa}; see that paper for details.}
    \label{fig:loc}
\end{figure*}

Experimentally, this prediction can be tested by extracting the entropy from measured charged hadron multiplicity distribution $P(N)$, 
\[
S_{\mathrm{DIS}} \simeq S_{\rm hadron}\;=\;-\sum_{N} P(N)\,\ln P(N)\,,
\]
and comparing it to the logarithm of the structure function, at different Bjorken $x$ and momentum transfer $Q^2$.
\smallskip
 
The validation of this MEL approach in DIS began \cite{Kharzeev:2021yyf,Hentschinski:2021aux} by comparing it to inclusive DIS data from H1 Collaboration at HERA~\cite{H1:2020zpd}. It was found that the contribution from sea quarks at finite energy was important although not dominant \cite{Kharzeev:2021yyf}, and incorporating both gluon and quark contributions \cite{Hentschinski:2021aux} improved the agreement between the data and the model. Building on these developments, the model was extended to describe another QCD process, diffractive DIS ~\cite{Hentschinski:2023izh}. Good agreement with the MEL model was observed~\cite{Hentschinski:2023izh}, as illustrated in Fig. \ref{fig:entropy_vsdata1}.
 \smallskip

To further test the MEL approach,  it is important to check whether it properly captures the QCD evolution. This was done by considering the data in varying rapidity windows \cite{Hentschinski:2024gaa}. For this purpose, we used 
the H1 $ep$ DIS data~\cite{H1:2020zpd} from HERA. The results are shown in Fig. \ref{fig:loc}; they imply that the MEL correctly captures the QCD evolution.
\smallskip

The existing HERA data thus already suggest the vailidity of MEL in DIS. Future measurements at the Electron Ion Collider will provide a decisive test over a wide kinematic range.

\subsection{MEL in jet fragmentation}

A complementary arena for testing maximal entanglement is provided by jet fragmentation in high-energy collisions, as discussed in section \ref{qsim}. Once a hard parton is produced, its subsequent fragmentation into hadrons involves the redistribution of quantum information among many degrees of freedom. If the underlying partonic state is maximally entangled, the reduced density matrix describing a subset of hadrons in a jet should again exhibit thermal-like features. In particular, multiplicity distributions and single-particle spectra in jets can be interpreted as arising from a reduced density matrix that is diagonal and approximately thermal.
\begin{figure*}[tbh]
    \centering
   \includegraphics[width=.9\textwidth]{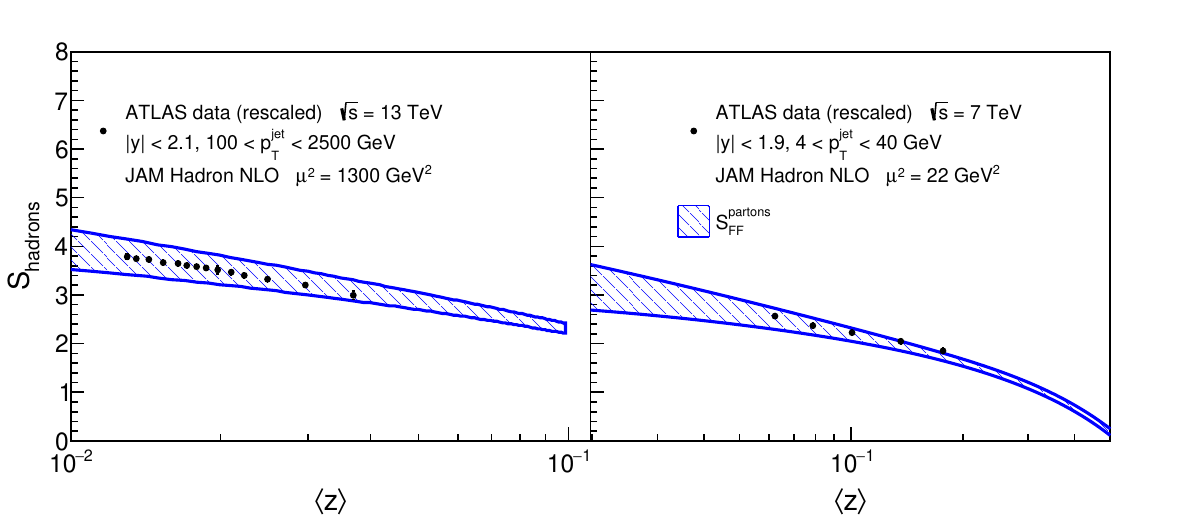}
\caption{The entropy $S_{\rm{hadrons}}$ as a function of $\left\langle z \right\rangle$ compared to $S_{\rm{FF}}^{\rm{partons}}$ -- incorporating gluons, $u$-(anti)quarks, and $d$-(anti)quarks -- is shown using JAM fragmentation functions at NLO for $\mu^2 = 1300$ GeV$^2$, compared with ATLAS data at $\sqrt{s} = 13$ TeV \cite{ATLAS:2019rqw} (left). Additionally, the results at $\mu^2 = 22$ GeV$^2$ are compared with ATLAS data at $\sqrt{s} = 7$ TeV \cite{ATLAS:2011eid} (right). The uncertainties are calculated at the 1$\sigma$ level. The total entropy $S_{\rm{FF}}^{\rm{partons}}$ is derived from the sum of the individual entropies of each parton, with each contribution normalized by the average fraction of jets produced by that parton from a PYTHIA simulation. From \cite{Datta:2024hpn}.}
\label{fig:ATLAS}
\end{figure*}

Recent analysis \cite{Datta:2024hpn} suggests that the entropy extracted from hadron multiplicities in jets exhibit MEL scaling behavior analogous to that observed in DIS, see Fig. \ref{fig:ATLAS}. In this figure, the hadron entropy $S_{\rm{hadrons}}$ computed from the measured multiplicity distribution is compared to the logarithm of the jet fragmentation function, as a function of variable $z=p^{\rm h}/P^{\rm jet}$ which is the ratio of hadron momentum to the momentum of the jet. 

Crossing symmetry relates the fragmentation functions and the parton distribution functions, so it is reassuring to see that MEL describes well both the jet production and DIS processes. From the present perspective, in both cases this is a direct consequence of tracing over unobserved degrees of freedom in a highly entangled many-body state. Jet fragmentation thus provides an important  test of entanglement-driven thermalization, extending the ideas discussed above from the initial state probed in DIS to the final state produced in high-energy collisions.
\smallskip

The MEL was also tested in inclusive proton-proton collisions at the LHC. In \cite{Kharzeev:2017qzs}, we have established that the cumulants computed from the geometric maximal entanglement multiplicity distribution \eqref{geomd} agree very well with the ones measured by the CMS Collaboration at the LHC \cite{CMS:2010qvf}. Furthermore, in \cite{Tu:2019ouv}, the MEL relation between the hadron entropy and the structure function was tested for inclusive $pp$ collisions at the LHC. It was found that MEL relation holds if one uses the data \cite{CMS:2010qvf}. It is also interesting that MEL relation is badly violated by the PYTHIA Monte-Carlo generator which is tuned to the data, but does not properly incorporate the entanglement \cite{Tu:2019ouv}. 
\smallskip

The origin of the apparent thermalization in high-energy collisions was  investigated in \cite{Baker:2017wtt} using the data of the ATLAS and CMS Collaborations at the LHC.
For this purpose, we analyzed the transverse momentum distributions in inclusive
inelastic pp collisions, single- and double-diffractive Drell-Yan production, and the production of the Higgs boson. 
We confirmed that the effective temperature of the produced hadrons is driven by the hard scattering scale, and found that this relation extends even to the Higgs boson production process.

\section{Is the Maximal Entanglement Limit universal?}

An important conceptual question is whether all quantum systems inevitably approach the Maximal Entanglement Limit (MEL), or whether this behavior is restricted to a particular class of dynamics. The likely answer is that MEL is not universal: its emergence requires sufficient dynamical complexity to render phase information operationally inaccessible, and this condition is not generically satisfied in all quantum systems.

In quantum chaotic systems, the approach to MEL is natural.  Generic pure states explore the accessible Hilbert space effectively, reduced density matrices approach thermal forms constrained only by conserved quantities, and entanglement entropy saturates near its maximal value. In this sense, chaotic dynamics provide a microscopic realization of MEL.

Integrable systems, by contrast, possess an extensive set of conserved quantities that severely restrict their dynamics. Phase correlations persist over long times, and the system fails to explore the full Hilbert space. Following a quench, such systems relax not to a thermal state but to a generalized Gibbs ensemble, with entanglement entropy that remains parametrically below the maximal value allowed by energy alone. Integrable systems therefore do not reach the MEL in the strict sense.
\smallskip

Nevertheless, exact chaos is not a necessary condition for the emergence of MEL. Effective maximal entanglement can arise whenever phase information becomes unobservable due to coarse graining or tracing over inaccessible degrees of freedom. In high--energy scattering, as we discussed in section \ref{opm}, Lorentz time dilation suppresses the observability of phases, while scale--invariant branching dynamics generate a rapidly growing number of entangled degrees of freedom. Under these conditions, even systems that are weakly coupled or integrable at the microscopic level may exhibit an effective approach to MEL.
\smallskip

Maximal entanglement should thus be viewed as an emergent property of systems in which dynamical complexity, coarse graining, and entanglement generation conspire to erase part of the information from observable quantities. The approach to MEL is not universal and strongly depends on the microscopic dynamics. Controlling the rate with which the system reaches the MEL, for example by tuning the parameters of the microscopic Hamiltonian or by intermediate measurements, is an important open problem.

\section{Summary and Future Directions}

In these lectures, we have developed the idea that the probabilistic structures of statistical mechanics and high--energy physics share a common origin in quantum entanglement and the geometry of Hilbert space.
Beginning with Boltzmann's original insight -- that thermodynamic behavior reflects the overwhelming typicality of microstates -- we discussed how this idea acquires a precise meaning in quantum theory. Following von Neumann and later work, we introduced the notion of typical quantum states in a small subsystem of a high--dimensional Hilbert space. We then saw how thermal behavior and probabilistic laws emerge naturally in the description of this subsystem from entanglement and the geomerty of Hilbert space, without any reference to classical randomness or ergodic assumptions.
\smallskip

We have then followed this logic into the relativistic quantum domain, where the relevant arena is a high-dimensional Hilbert space on the light cone. In this setting, the familiar probabilistic parton model emerges naturally as a reduced description of an underlying pure quantum state, once unobservable phases of Fock states are traced out. 
\smallskip

The emergence of a probabilistic description is not restricted to weak--coupling or perturbative regimes. More generally, it arises whenever a subset of degrees of freedom becomes inaccessible to measurement and must be traced over.  As emphasized by Dirac in his classification of relativistic dynamics \cite{Dirac:1949cp}, the light--cone (or ``front form'') is distinguished by the fact that the number of kinematical generators of the Poincar\'e group -- that is, generators that do not involve the Hamiltonian -- increases to seven, compared to six in the conventional instant form. This enlargement of the kinematical symmetry implies an additional freedom in specifying the orientation of the quantization surface, which is not fixed by the dynamics.
\vskip0.2cm

It is therefore natural to conjecture \cite{Kharzeev:2021nzh} that the effective Haar scrambling of the reduced density matrix discussed above can be understood as an averaging over this additional kinematical degree of freedom inherent to light--cone quantization. From this viewpoint, tracing over unobserved degrees of freedom corresponds to integrating over the orientations associated with the extra kinematical symmetry, with the invariant Haar measure of the corresponding group. This perspective suggests that the probabilistic structures characteristic of high--energy scattering are rooted not in perturbative dynamics, but in the geometric and group--theoretic properties of relativistic quantum theory formulated on the light cone. It will be interesting to test and develop this idea in the future.
\smallskip

Another central message in these lectures has been that {\bf high-energy experiments} provide an exceptional laboratory for testing foundational ideas in quantum statistical physics. The diagonal density matrix underlying the parton model, the thermal-like hadron abundances and transverse momentum distributions, fast thermalization, and hadron multiplicity distributions in deep inelastic scattering and jet fragmentation can all be understood as manifestations of entanglement-driven typicality. 
\smallskip

The MEL relation \cite{Kharzeev:2017qzs} between the hadron entropy and parton structure or fragmentation functions, mirroring Boltzmann's $S=\ln\Omega$, quantifies this viewpoint and makes it amenable to direct experimental tests. As discussed in these lectures, the early results are encouraging, but this effort should definitely continue. 
A promising new direction is to probe quantum entanglement in the fragmentation of high--energy jets using observables borrowed from quantum information theory, such as mutual information. In this approach, an individual jet, or a part of a jet (defined, for example, by angular cones, momentum fractions, or rapidity intervals) is  treated as a subsystem of a single quantum state produced in the hard scattering. The mutual information,
$$
I(A:B) = S_A + S_B - S_{A\cup B},
$$
quantifies the total amount of correlation between two such subsystems, including correlations of purely quantum origin. Unlike traditional correlation functions, mutual information is insensitive to the choice of basis and captures both classical and quantum correlations in a unified manner.
\smallskip


Let us sketch how the extraction of entanglement entropy from hadron multiplicity distributions in a single jet can be generalized to access mutual information in events with two well--separated jets. For simplicity let us consider $e^+e^-\to q\bar q$ annihilation, where the color structure is simpler; however this approach can be generalized to jet production in DIS or hadron collisions. 
\smallskip

 The two jets define subsystems $A$ and $B$, consisting of the hadrons clustered into jet~1 and jet~2, respectively. In direct analogy with the single--jet analysis, we define the hadron multiplicities $n_1$ and $n_2$ in the two jets (with a fixed transverse--momentum threshold), and measure the joint probability distribution
$$
P(n_1,n_2) \, ,
$$
together with the individual (marginal) distributions $P_1(n_1)$ and $P_2(n_2)$.
\smallskip

In the high-energy regime, where the phases are unobservable and the reduced density matrices are effectively diagonal in the occupation--number basis, the von Neumann entropies are well approximated by Shannon entropies. The mutual information between the two jets then reduces to
$$
I(A\!:\!B)
\simeq
\sum_{n_1,n_2} P(n_1,n_2)\,
\ln\!\left[
\frac{P(n_1,n_2)}{P_1(n_1)\,P_2(n_2)}
\right]
\;\ge\; 0 .
$$
This quantity measures the entanglement between the two jets. By construction, $I(A\!:\!B)$ vanishes if the two jets fragment independently (as is traditionally assumed in perturbative QCD) and is positive whenever their fragmentation histories remain correlated. 
\smallskip

The quantum simulation in Schwinger model performed in \cite{Florio:2023dke} suggests that the entanglement between the produced hadrons is substantial when the rapidities at which the hadron multiplicity distributions in $n_1(\eta_1)$ and  $n_2(\eta_2)$ are measured, possess rapidity difference $\Delta \eta \leq 2$. If high--energy evolution drives the system toward MEL, one expects the joint distribution $P(n_1,n_2)$ to approach a universal form determined by the average multiplicities, with the mutual information exhibiting the corresponding scaling.
\vskip0.2cm

This analysis can be systematically refined by replacing multiplicity with a richer set of observables defining each subsystem, such as binned energy flow within the jet, angular substructure, or groomed observables. In all cases, the experimental task is to measure the corresponding joint probability distributions and construct the mutual information. Present (RHIC, LHC) and future EIC data thus offer an opportunity to probe entanglement and information flow in jet fragmentation.
\smallskip


The Maximal Entanglement Limit provides a conceptual framework that opens a number of concrete directions for future {\bf theoretical research}. One promising avenue is the use of quantum--information--theoretic measures to distinguish chaotic from integrable dynamics by checking whether a given Hamiltonian efficiently explores its accessible Hilbert space or remains constrained by integrability. 
As an intriguing recent example of work in this direction, let me mention the paper \cite{Sharipov:2024lah} that utilizes a metric tensor on manifolds of quantum states\footnote{This Riemannian quantum metric was originally introduced by D. Hill and J.A.~Wheeler \cite{Hill:1952jb} to describe collective nuclear dynamics in fission.}  \cite{Provost:1980nc} to argue that ergodic quantum systems are described by smooth manifolds, whereas integrability exhibits itself as a conical singularity.
Developing a systematic classification of quantum field theories and many--body systems based on information--theoretic criteria would be very valuable.

A second direction concerns the formulation of a general theory for the approach to the Maximal Entanglement Limit grounded in the geometry of Hilbert space. From this viewpoint, unitary time evolution corresponds to trajectories on a high--dimensional projective manifold, while MEL emerges as a consequence of concentration of measure in exponentially large energy shells. Understanding how curvature, dimensionality, and symmetry constraints of Hilbert space control entanglement generation could lead to a geometric classification of thermalization pathways, clarifying when and how typicality sets in. 

A closely related problem is how to control, accelerate, or inhibit the approach to MEL. On the theoretical side, this involves tuning Hamiltonian parameters to interpolate between integrable and chaotic regimes, as well as introducing controlled perturbations that lift degeneracies or enhance mixing in Hilbert space. Equally important is the role of intermediate, or mid--circuit, measurements, which can partially collapse the wave function and compete with entanglement growth. Understanding the interplay between unitary evolution, measurements, and entanglement production is essential for mapping out phase diagrams of quantum dynamics.  

In QCD, an important direction is the extension of the parton model to explicitly incorporate light--cone zero modes and the associated degeneracies in light--cone energies. As discussed in these lectures, such degeneracies preserve phase coherence and invalidate a purely probabilistic description. A systematic framework that treats zero modes on the same footing as ordinary partonic degrees of freedom would be a significant step toward describing nonperturbative QCD vacuum structure within a light--cone formalism. 
\smallskip

These theoretical developments are not of purely conceptual interest. Control over entanglement growth and approach to MEL is directly relevant for {\bf practical applications} across multiple fields. In quantum information science, excessive entanglement  to the environment is closely tied to qubit decoherence and information loss, while insufficient entanglement among qubits limits computational power. In quantum Machine Learning, rapid convergence to MEL under generic circuits provides a natural explanation for the ``barren plateau problem" \cite{Marrero:2021xmi}, where gradients vanish due to concentration of measure. In strong--field and laser--matter interactions, understanding how quickly quantum systems approach maximal entropy regime for a particular pulse shape (or quench protocol) is essential for controlling energy deposition, ionization cascades, and laser-induced nuclear processes.
\smallskip

Viewed from this broader perspective, the Maximal Entanglement Limit is not merely an organizing principle for statistical and high--energy physics, but a bridge between fundamental science and pressing challenges in quantum technologies. Developing a quantitative, geometry--based theory of MEL and its controlled violation thus represents a promising and interdisciplinary frontier for future research.

\section*{Acknowledgements}

I am very grateful to Micha\l{} Prasza\l{}owicz and his colleagues for warm hospitality in Zakopane. 
I thank S. Forte, R. Janik, G. Korchemsky, Yu. Kovchegov, L. McLerran, S. Mr\'owczy\'nski, M. Prasza\l{}owicz, S. Reddy and other participants of the School for an inspiring atmosphere, and for stimulating questions and discussions. 
\smallskip

I am deeply indebted to my collaborators for many years of close collaboration through which these ideas took shape:
K.~Baker, P.~Castorina, J.~Datta, A.~Deshpande, Yu. Dokshitzer, A. Florio, D. Frenklakh, A.~Gorsky, S.~Grieninger, U.~G\"ursoy, K.~Hao, M. Hentschinski, K.~Ikeda, V. Korepin, K. Kutak, E.~Levin, F. Loshaj, E.~Marroquin, L.~McLerran, C.~Na\"im, A. Palermo, V. Pascuzzi, J.~Pedraza, K.~Rajagopal, G. Rossi, H.~Satz, S.~Shi, C.~Trallero, K. Tu, K. Tuchin, T. Ullrich, G. Veneziano, R.~Venugopalan, N. Wiebe, K. Yu, I.~Zahed, H.-U. Yee and K. Zhang.
\smallskip

This work was supported by the U.S. Department of Energy, Office of Science, Office of Nuclear Physics under contract No.DE-FG88ER41450 and by the U.S. Department of Energy, Office of Science, National Quantum Information Science Research Centers, Co-design
Center for Quantum Advantage (C2QA) under Contract No.DE-SC0012704. Part of the work on these lectures was performed at the Aspen Center for Physics, supported by National Science Foundation grant PHY-2210452, during the 2025 Summer program on ``Strongly interacting quantum matter at the Electron Ion Collider" organized by Z. Meziani, F.~Salazar, Y. Zhao and D.K.


\end{document}